\documentclass[twocolumn,showpacs,preprintnumbers,amsmath,amssymb]{revtex4}
\usepackage{graphicx}
\usepackage{psfrag}
\usepackage{dcolumn}
\usepackage{bm}
\usepackage{amsthm}
\usepackage[all]{xy}

\newcommand{\ke}{\kern0.2em}
\newcommand{\ts}{\kern0.1em}

\newcommand{\from}{\colon}
\newcommand{\too}{\longrightarrow}
\newcommand{\mapstoo}{\longmapsto}
\newcommand{\inv}{^{-1}}
\newcommand{\q}{\quad}
\newcommand{\ca}[1]{{\cal#1}}
\newcommand{\bb}[1]{\mathbb#1}
\newcommand{\eu}[1]{{\mathfrak#1}}
\newcommand{\sm}{\!\setminus\!}
\newcommand{\im}{\mathop{\rm image}\nolimits}
\newcommand{\supp}{\mathop{\rm supp}\nolimits}
\newcommand{\pr}{\mathop{\rm pr}\nolimits}
\newcommand{\id}{\mathop{\rm id}\nolimits}
\newcommand{\Hom}{\mathop{\rm Hom}\nolimits}
\newcommand{\Aut}{\mathop{\rm Aut}\nolimits}
\newcommand{\chr}{\mathop{\rm char}\nolimits}
\newcommand{\by}{\!\times\!}
\newcommand{\tschech}{\check{\phantom{\vbox{\hrule width4pt height4.7pt}}}}

\renewcommand{\phi}{\varphi}

\newtheorem{thm}{Theorem}
\newtheorem{prop}[thm]{Proposition}
\newtheorem{la}[thm]{Lemma}
\newtheorem{cor}[thm]{Corollary}

\theoremstyle{remark}
\newtheorem*{ex}{Example}
\newtheorem*{exs}{Examples}

\begin{document}
\title{Homological invariants of stabilizer states}
\author{Klaus Wirthm\"uller}
\email{wirthm@mathematik.uni-kl.de}
\affiliation{Fachbereich Mathematik der Technischen Universit\"at Kaiserslautern}
\date{November 12, 2006}
\begin{abstract}
We propose a new kind of invariant of multi-party stabilizer states with respect to local Clifford equivalence. These homological invariants are discrete entities defined in terms of the entanglement a state enjoys with respect to arbitrary groupings of the parties, and they may be thought of as reflecting entanglement in a qualitative way. We investigate basic properties of the invariants and link them with known results on the extraction of GHZ states.
\end{abstract}
\maketitle
\section{Introduction}
\subsection{Overview}
In quantum information theory states that are jointly held by several parties play a central role, and are at the heart of the notion of entanglement. Correspondingly the problem of classifying such states has attracted a lot of interest, the correct notion of equivalence being that of local unitary (LU) equivalence of states, which allows each party to apply arbitrary local unitary operators. While classification has been successful in special cases involving few parties it is quite unreasonable to expect a meaningful answer to the classification problem in general.

Among all quantum states the so-called stabilizer states form a subclass which is easier to investigate since much of the multilinear structure of stabilizer states is reflected in the additive structure of the corresponding stabilizer group. On the other hand many important states like GHZ states can be realised as stabilizer states, and furthermore it has been found \cite{van_den_nest3} that LU equivalence between sufficiently entangled stabilizer states tends to coincide with local Clifford (LC) equivalence, a fact that by and large reduces the classification of stabilizer states to that of stabilizer groups. Though conceptually easier the latter problem still is far too general to have a useful answer. It therefore appears reasonable to look for and study invariants of states which are intrinsically coarse in the sense that many non-equivalent states are expected to share the same invariants --- provided of course that at least some essential features of a state are still reflected in its invariants.

In this paper we propose and construct such a set of invariants. In the case we have in mind as the most important one we consider a finite set $P$ of parties, each of which controls a state space $\ca H_p$ with some (finite) number of qubits. To each pure stabilizer state $|\psi\rangle$ we will assign a sequence of finite dimensional vector spaces $H^0(|\psi\rangle),\dots,H^{|P|}(|\psi\rangle)$ which are invariant with respect to LC equivalences acting on $|\psi\rangle$, and also with respect to permutations of the parties. Our first aim in the construction of these invariants is to capture the entanglement that is present in $|\psi\rangle$ with respect to all possible subdivisions of $P$. It appears that the notion of sheaf, which was introduced and developed in great generality by the French school of algebraic geometers since the 1940s, is perfectly suited to accommodate the desired data in a single object, which we call a partition sheaf. In a second step we intentionally reduce the amount of information by applying cohomology to that sheaf --- a process whose formal properties are well understood and which in algebraic geometry and other mathematical contexts has proven to preserve essential, and erase superficial information.

By its very nature the homological invariants of a state are a qualitative rather than quantitative measure of entanglement, and more specifically of entanglement expressible by linear data (rather than multilinear like the Schmidt measure introduced in \cite{eisert2}). Homological invariants are easily and efficiently computed from their definition in terms of standard linear algebra. Also, their behaviour under standard processes like coarsening the distribution of the quantum system among the parties, or joining quantum systems is, at least in part, computable.

A closer investigation of homological invariants reveals the presence of a duality which eventually comes from the symplectic structure on the stabilizer groups, and the principal part of Section \ref{prod_section} is dedicated to the investigation of this duality. 

For many states we have explicitly calculated the homological invariants, and we present some of the results in Section \ref{ex_appl_section}. Interestingly, duality of homological invariants seems to be an important mathematical theme underlying the work of \cite{bravyi3} on extraction of GHZ states, where it has appeared in an ad hoc way in some of the proofs. We therefore show how to recover and complement results of the cited work, giving proofs which are conceptual in terms of the invariants.
\subsection{Stabilizer states}
The stabilizer formalism was introduced in \cite{gottesman_96} in the context of 
error-correcting quantum codes, and since has been explained and widely used in varying degree of generality \cite{calderbank4,knill,rains,gottesman_98,hostens3}. The purpose of the following very brief account is to fix the terminology and notation used below.

The notion of stabilizer state is based on the mathematical one of Heisenberg group, see \cite{mumford}. Here we consider but a restricted class of such groups, built from data comprising a finite dimensional vector space $G$ over a finite field $\bb F=\bb F_p=\bb Z/p$ of prime order, and a symplectic form
$\omega\from G\otimes G\to\bb F$ --- that is, a skew-symmetric bilinear $\omega$ that induces an isomorphism
$G\owns g\mapsto\omega(g,?)\in G\tschech=\Hom_\bb F(G,\bb F)$. The corresponding Heisenberg group is the central extension
\begin{displaymath}
  \begin{xy}
    \xymatrix{{1}\ar@{->}[r]&
                {\rule[-2.4pt]{0pt}{10pt}S^1}\ar@{->}[r]&
                {\rule[-2.4pt]{0pt}{10pt}\tilde G}\ar@{->}[r]&
                {G}\ar@{->}[r]&
                {0}\\}
  \end{xy}
\end{displaymath}
of the multiplicative group $S^1=\left\{\lambda\in\bb C\,\big|\,|\lambda|=1\right\}$ by the additive group $G$ determined by
\begin{displaymath}
  G\times G\owns(g,h)\mapstoo e(g,h)=\exp2\pi i\,\omega(g,h)/p\in S^1
\end{displaymath}
in the sense that $e(g,h)=\tilde g\,\tilde h\,{\tilde g}\inv{\tilde h}\inv$ is the commutator of any two group elements representing $g$ and $h$. By the theorem of Stone, von Neumann, and Mackey (SNM) every Heisenberg group has a unique irreducible unitary representation space $\ca H$ on which the subgroup $S^1$ acts by scalars in the natural way. In the basic case where $p=2$ and
$G=\bb F_2\times\bb F_2$ carries the canonical symplectic form this representation may be realised sending the nonzero vectors of $G$ to the Pauli matrices $\sigma_x$, $\sigma_y$, and $\sigma_z$ with their action on one qubit. More generally, taking $G$ as a finite orthogonal sum of $l$ copies of $\bb F_2\times\bb F_2$ the resulting SNM representation $\ca H$ is an $l$-qubit space with one set of Pauli operators acting on each qubit. Therefore physicists usually consider the Heisenberg group $\tilde G$ as a subgroup of the unitary group $U(\ca H)$ and call it a Pauli group.

Let now $L\subset G$ be an isotropic subspace\ts: $\omega(g,h)=0$ for all $g,h\in L$. Then the natural projection $\tilde G\to G$ admits a section $s\from L\to\tilde G$ over $L$, so that $s(L)\subset\tilde G$ is an abelian subgroup. Its fixed space
\begin{displaymath}
  \ca H^{s(L)}=\left\{|\psi\rangle\in\ca H\,\big|\,
    g|\psi\rangle=\psi\rangle\mbox{ for all }g\in s(L)\right\}
\end{displaymath}
is a stabilizer state corresponding to $L$. It is a vector subspace of $\ca H$ of dimension $2^{\dim G-2\dim L}$, and therefore a mixed state in general while a pure state occurs if and only if $L\subset G$ has the maximal dimension $\frac{1}{2}\cdot\dim G$\ts: such $L$ are called lagrangian subspaces of $G$.

Automorphisms of the Heisenberg group $\tilde G$ are --- by definition --- required to restrict to the identity on $S^1$, and they form a group $\Aut\tilde G$ which is an extension of the dual $\Hom(G,S^1)=G\tschech$ by the group $Sp(G)$ of symplectic automorphisms of $G$. Every element of $\Aut\tilde G$ gives rise to a unitary transformation of $\ca H$, called a Clifford automorphism. As to the classification of stabilizer states, the automorphisms from $G\tschech$ just permute the different sections $s$ over $L\subset G$, so that the Clifford classification of states reduces to the symplectic classification of the isotropic subspaces $L\subset G$. An interesting classification problem only arises if some additional structure is given, like a finite direct decomposition of $G=\bigoplus_{p\in P}G_p$ which is orthogonal with respect to $\omega$. Then the local symplectic group
\begin{displaymath}
  \prod_{p\in P}Sp(G_p)\subset Sp(G)
\end{displaymath}
acts on $G$ in such a way as to restrict each party's control to the subspace owned by it, and the analogously defined local Clifford equivalence of stabilizer states in $\ca H=\bigotimes_{p\in P}\ca H_p$ corresponds to exactly this local symplectic equivalence of isotropic subspaces.

For the mere construction of homological invariants no explicit reference to stabilizer states is needed nor even appropriate since the vector space $G$, its direct sum decomposition $G=\bigoplus_{p\in P}G_p$, and the subspace $L\subset G$ is all the data required. Neither any assumption on the base field $\bb F$ will be made, and the symplectic form $\omega$ only comes in at a later stage.
\section{The construction}
\subsection{Partition spaces}
Let $P$ be a finite set\ts; the elements of $P$ will be referred to as \textit{parties}. In this subsection we will assign to $P$ its \textit{partition space} $X=X(P)$. The partition space is a topological space that violates the Hausdorff separation axiom since distinct points of $X$ cannot be separated by neighbourhoods, and one should think of it not as a geometric but rather a formal object that in spite of its simplicity serves as a useful organizing tool. In fact it is simply modelled on the two point topological space $\{0,1\}$ in which the point $0$, or more precisely the subset $\{0\}$ is closed while $\{1\}$ is not a closed subset of $X$\ts: thus $\emptyset$, $\{1\}$, $X$ is the complete collection of open subsets of $X$. The underlying idea, borrowed from algebraic geometry, is to think of $0$ as the unique \textit{special}, and of $1$ as the \textit{general} point of $\{0,1\}$ --- though we do not require these
notions in a formal way.

\begin{figure}
  \psfrag{1}[c][c]{$1$}
  \psfrag{2}[c][c]{$2$}
  \psfrag{3}[c][c]{$3$}
  \psfrag{U1}[c][c]{$U_1$}
  \psfrag{U2}[c][c]{$U_2$}
  \psfrag{X2}[c][c]{$X_2$}
  \psfrag{X3}[c][c]{$X_3$}
  \includegraphics{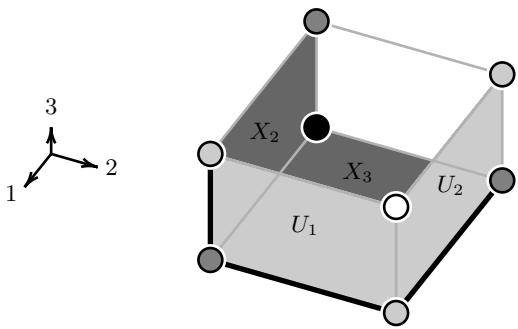}
  \caption{\label{loc_closed}
    The locally closed set $(U_1\cup U_2)\cap(X_2\cup X_3)$
    in the three party partition space $X$ has the canonical decomposition
    $(U_1\cap X_2)\cup(U_1\cap X_3)\cup(U_2\cap X_3)$. The black lines
    indicating this set as well as the walls indicating $U_1$ etc.\ are, of
    course, imaginary since in reality $X$ comprises just eight points.}
\end{figure}

The partition space
\begin{displaymath}
  X=X(P)=\{0,1\}^P
\end{displaymath}
is defined as the cartesian power of the topological space $\{0,1\}$ just described, thus an element of $X$ is a word ${(x_p)}_{p\in P}$ whose letters $x_p$ are zero or one and are labelled by $p\in P$. For a subset of parties $S\subset P$ we write
\begin{eqnarray*}
X_S &= &\left\{x\in X\,|\,x_p=0\mbox{ for all }p\in S\right\}\mbox{ and}\\
U_S &= &\left\{x\in X\,|\,x_p=1\mbox{ for all }p\in S\right\},
\end{eqnarray*}
while $X_p$, $U_p$ will be shorthand for $X_{\{p\}}$ and $U_{\{p\}}$. For any $S$ these are closed respectively open subsets of $X$\ts; indeed $X_S$ is the smallest closed set that contains the point $x\in X$ with $x_p\!=\!0\;(p\!\in\!S)$ and $x_p\!=\!1\;(p\!\notin\!S)$ whereas $U_S$ is the smallest open set that contains the complementary point $y\in X$ with $y_p=1\;(p\in S)$ and $y_p=0\;(p\notin S)$. Every closed subset $Z\subset X$ may be written as a union
\begin{displaymath}
  Z=X_{S_1}\cup\cdots\cup X_{S_r}\q(r\in\bb N),
\end{displaymath}
and imposing the condition
\begin{displaymath}
  S_i\not\subset S_j\mbox{\q for }i\neq j
\end{displaymath}
forces this decomposition to be that into the irreducible components of $Z$\ts; in particular it then is unique up to the order of the summands. In view of the symmetry $0\leftrightarrow1$ we obtain an analogous representation
\begin{displaymath}
  U=U_{S_1}\cup\cdots\cup U_{S_r}
\end{displaymath}
of an arbitrary open set $U\subset X$. In particular the sets $U_S$ form a basis of the topology of $X$. Note the special cases
\begin{displaymath}
  X_\emptyset=U_\emptyset=X\mbox{ and }X_P=\{\bullet\},\;U_P=\{\circ\},
\end{displaymath}
the latter sets comprising the unique closed, respectively open point of $X$. More generally, every locally closed subset of $X$, say $U\cap Z$ with open $U=U_{S_1}\cup\cdots\cup U_{S_r}$ and closed $Z=X_{T_1}\cup\cdots\cup X_{T_s}$ has a canonical decomposition
\begin{displaymath}
  U\cap Z=\bigcup_{\{(\sigma,\tau)\,|\,S_\sigma\cap T_\tau=\emptyset\}}
    U_{S_\sigma}\cap X_{T_\tau}
\end{displaymath}
(though an arbitrary union of this form need not be locally closed), see Fig.~\ref{loc_closed}.
Finally note the canonical homeomorphisms
\begin{displaymath}
  X_S\approx X(P\sm S)\approx U_S
\end{displaymath}
for all $S\subset P$.
\subsection{Sheaves on partition spaces}
A presheaf $\ca F$ on a topological space $X$ assigns to each open subset $U\subset X$ a set $\ca F(U)$, whose elements are called sections over $U$, and to each inclusion of open subsets $U\subset V$ a mapping $\ca F_{UV}\from\ca F(V)\to\ca F(U)$ called the restriction such that the relation $\ca F_{UV}\circ\ca F_{VW}=\ca F_{UW}$ holds whenever $U\subset V\subset W$. Presheaves often carry additional algebraic structures, and all presheaves considered here will be presheaves of vector spaces over a fixed field $\bb F$, that is, all $\ca F(U)$ will be vector spaces and all restriction maps, linear homomorphisms.

A sheaf on $X$ is a presheaf that satisfies two extra conditions\ts:
\begin{itemize}
  \item If $U\subset X$ is open, $U=\bigcup_{i\in I}V_i$ a covering of $U$ by
    open sets, and $f\in\ca F(U)$ then $\ca F_{V_iU}(f)=0$ for all $i\in
    I$ implies $f=0$.
  \item If $U\subset X$ is open, $U=\bigcup_{i\in I}V_i$ a covering of $U$ by
    open sets, and $f_i\in\ca F(V_i)$ are given such that
    $\ca F_{V_i\cap V_j,V_i}=\ca F_{V_i\cap V_j,V_j}=0$ for all $i,j\in I$ then
    there exists a section $f\in\ca F(U)$ with $\ca F_{U\cap V_iU}(f)=f_i$ for
    all $i\in I$.
\end{itemize}
Thus the property that characterises sheaves among presheaves is that sections over $U$ may be described by local sections that are compatible on intersections. From every presheaf $\ca F$ on $X$ a well-defined sheaf may be constructed essentially by enforcing the two extra conditions.

Rather than by their spaces of sections sheaves may also be characterised by their stalks. In general the stalk of the sheaf $\ca F$ at the point $x\in X$ is defined as $\lim_U\ca F(U)$ where the direct limit is taken over the system of all open neighbourhoods of $x$ in $X$. For the partition spaces $X=X(P)$ that we consider here this system contains the set $U_{\{p\in P\,|\,x_p=1\}}$ as its greatest element (that is, smallest with respect to inclusion of sets): thus forming the direct limit is perfectly unnecessary, and the stalk $\ca F(x)=\ca F(U_{\{p\in P\,|\,x_p=1\}})$ reduces to a particular space of sections of $\ca F$.

The standard notions of linear algebra also make sense for sheaves if applied to the latters' stalks. In particular there are the notions of direct sum, of sub and quotient sheaves, of sheaf homomorphisms giving rise to kernel and image subsheaves, and of exact sequences
\begin{displaymath}
  \begin{xy}
    \xymatrix{{\cdots}\ar@{->}[r]&
                {\ca E}\ar@{->}[r]^f&
                {\ca F}\ar@{->}[r]^g&
                {\ca G}\ar@{->}[r]&
                {\cdots}\\}
  \end{xy}
\end{displaymath}
of sheaves on $X$ (exactness at $\ca F$ meaning that $\im f=\ker g$). A short exact sequence
\begin{displaymath}
  \begin{xy}
    \xymatrix{{0}\ar@{->}[r]&
                {\ca E}\ar@{->}[r]&
                {\ca F}\ar@{->}[r]&
                {\ca G}\ar@{->}[r]&
                {0}\\}
  \end{xy}
\end{displaymath}
is a means of stating that $\ca F$ is an extension of $\ca E$ by $\ca G$.

The simplest presheaves on a topological space $X$ are constant and assign to each open $U\subset X$ the same vector space, say $L$. By abuse of language the corresponding constant sheaf is also written $L$\ts; to every open $U\subset X$ it assigns the cartesian product $L^{\pi(U)}=\prod_{\pi(U)}L$ where $\pi(U)$ is the set of connected components of $U$. In particular one has $L(U)=L$ for all non-empty connected open $U\subset X$ while $L(\emptyset)=0$ is the zero space.

More useful are the sheaves on $X$ which we call \textit{elementary}. Such a sheaf is based on data comprising a vector space $L$ and an open subset $V\subset X$. Writing the inclusion as $j\from V\hookrightarrow X$ the corresponding elementary sheaf is $j_!L$, the extension of the constant sheaf $L$ from $V$ to $X$ by zero. It has the properties
\begin{displaymath}
  j_!L(U)=
    \left\lbrace
      \begin{array}{cl}
        L &\mbox{ if }U\mbox{ is non-empty connected and }U\subset V\\
        0 &\mbox{ if }U\mbox{ is connected and }U\not\subset V.
      \end{array}
    \right.
\end{displaymath}
By a \textit{quasi-elementary} sheaf on $X$ we mean one that can be obtained by
successive extension of elementary sheaves.

Most important in our context are sheaves of the following kind. Suppose that we are given a set $P$ of parties and for each party $p\in P$ a finite dimensional vector space $G_p$ over the fixed field $\bb F$. We write
\begin{displaymath}
  G_S:=\bigoplus_{p\in S}G_p
\end{displaymath}
for each subset $S\subset P$ and put $G:=G_P$. Let $L\subset G$ be a linear subspace. For each non-empty open subset $U$ of the partition space $X=X(P)$ we put
\begin{displaymath}
  \ca FL(U)=L\cap G_{\{p\in P\,|\,U\subset U_p\}}\ts;
\end{displaymath}
these spaces clearly define a presheaf $\ca FL$ on $X$ whose restriction homomorphism in case $U\subset V$ is just the inclusion
\begin{displaymath}
  L\cap G_{\{p\in P\,|\,V\subset U_p\}}\hookrightarrow
  L\cap G_{\{p\in P\,|\,U\subset U_p\}}.
\end{displaymath}
Since every non-empty open $U\subset X$ is connected $\ca FL$ in fact is a sheaf on $X$, and its stalk at $x\in X$ is
\begin{displaymath}
  \ca FL(x)=\ca FL(U_S)\mbox{ with }S=\{p\in P\,|\,x_p=1\}.
\end{displaymath}
We will call $\ca FL$ the \textit{partition sheaf} defined by $P$ and $L\subset G$. If $U$ is explicitly given as $U=U_{S_1}\cup\cdots\cup U_{S_r}$ then
\begin{displaymath}
  U\subset U_p\iff p\in S_i\mbox{ for all }i,
\end{displaymath}
so that $\ca FL(U)=L\cap G_{S_1\cap\cdots\cap S_r}$. Note in particular
$\ca FL(U_S)=L\cap G_S$ and the extreme cases
\begin{displaymath}
  \ca FL\{\circ\}=\ca FL(U_P)=L\mbox{ and }\ca FL(X)=\ca FL(U_\emptyset)=0.
\end{displaymath}

If $L'\subset L$ is a linear subspace then $\ca FL'\subset\ca FL$ is a subsheaf. We also may apply exterior powers to $\ca FL$, and have
\begin{displaymath}
  \begin{array}{cc}
    \Lambda^k\ca FL(U) &=\Lambda^k\left(L\cap G_{\{p\in P\,|\,U\subset U_p\}}\right)\\
                       &=\Lambda^kL\cap\Lambda^kG_{\{p\in P\,|\,U\subset U_p\}}     
  \end{array}
\end{displaymath}
for all $k\in\bb N$.
\begin{prop}\label{is_quasi-elementary}
For all subspaces $L'\subset L\subset G$ and all $k\in\bb N$ the quotient sheaf $\Lambda^k\ca FL/\Lambda^k\ca FL'$ is a quasi-elementary sheaf.
\end{prop}
\begin{proof}
We argue by induction on the dimension of $L/L'$. The assertion being trivial in case $L'=L$ we assume that $L'\subset L$ is a proper subspace. Since extensions of quasi-elementary sheaves clearly inherit that property we may further suppose that $\dim L=\dim L'+1$. Define the sheaf $\ca C$ as the cokernel of the short exact sequence
\begin{displaymath}
  \begin{xy}
    \xymatrix{{0}\ar@{->}[r]&
                {\ca FL'}\ar@{->}[r]&
                {\ca FL}\ar@{->}[r]&
                {\ca C}\ar@{->}[r]&
                {0.}\\}
  \end{xy}
\end{displaymath}
For any vector $g\in G$ we let $\supp g:=\left\{p\in P\,|\,g_p\ne0\right\}$ be the \textit{support} of $g$. We put
\begin{displaymath}
  W=\bigcup_{g\in L\setminus L'}U_{\supp g}\subset X
\end{displaymath}
and claim that $\ca C\simeq j_!(L/L')$ where $j\from W\hookrightarrow X$ is the open inclusion.

It suffices to study sections in $\ca C(U_S)$ for $S\subset P$. If $U_S\subset W$ then there is some $g\in L\sm L'$ with $U_S\subset U_{\supp g}$, so that $\supp g\subset S$. Therefore we have $g\in L\cap G_S=\ca FL(U_S)$, hence
\begin{displaymath}
  \ca FL(U_S)=\ca FL'(U_S)\oplus\langle g\rangle
\end{displaymath}
in this case.

If on the other hand $U_S\not\subset W$ then we pick some $x\in U_S\sm W$. Thus we have $x\in U_S$ but for all $g\in L\sm L'$ necessarily $x\notin U_{\supp g}$ and therefore $\supp g\not\subset S$. We conclude
\begin{displaymath}
  \ca FL(U_S)=\ca FL'(U_S)
\end{displaymath}
in this second case.

We thereby have shown that the canonical sheaf homomorphism $\ca FL/\ca FL'\to L/L'$ induces an isomorphism $\ca C\simeq j_!(L/L')$, which establishes our claim. The conclusion of the proposition now follows from the identity
\begin{displaymath}
  \Lambda^k\ca FL/\Lambda^k\ca FL'=\Lambda^{k-1}\ca FL'\;\otimes\;\ca FL/\ca
  FL',
\end{displaymath}
the fact that the tensor product of elementary sheaves is elementary, and induction on $k$.
\end{proof}
\subsection{Cohomology}
Let $\ca F$ be a sheaf of vector spaces on a topological space $X$. Associated with any such sheaf is the sequence $H^0(X;\ca F)$, $H^1(X;\ca F)$ \dots of cohomology groups (traditionally referred to as groups even if they are, in fact, vector spaces over the same field as $\ca F$)\cite{godement,hartshorne}. It is often convenient to combine all groups from the sequence into the single graded vector space
\begin{displaymath}
  H^\ast(X;\ca F):=\bigoplus_{i=0}^\infty H^i(X;\ca F).
\end{displaymath}
Apart from $H^0(X;\ca F)=\ca F(X)$, the space of global sections of $\ca F$, these groups have no simple direct interpretation and in no case of interest it is possible to compute them directly. Nevertheless in many instances cohomology of sheaves is highly computable by methods that rely on the formal properties of cohomology rather than one particular definition. Among the most prominent of these properties is the fact that every homomorphism $\ca F\to\ca G$ of sheaves over $X$ induces linear mappings $H^i(X;\ca F)\to H^i(X;\ca G)$ in a functorial way (that is, compatible with composition of maps), and that every exact sequence of sheaves
\begin{displaymath}
  \begin{xy}
    \xymatrix{{0}\ar@{->}[r]&
                {\ca E}\ar@{->}[r]&
                {\ca F}\ar@{->}[r]&
                {\ca G}\ar@{->}[r]&
                {0}\\}
  \end{xy}
\end{displaymath}
gives rise to a long exact cohomology sequence
\begin{displaymath}
  \begin{xy}
    \xymatrix@R0.4pc{{\rule[-2.4pt]{0pt}{10pt}0}\ar@{->}[r]&
                      {H^0(\ca E)}\ar@{->}[r]&
                      {H^0(\ca F)}\ar@{->}[r]&
                      {H^0(\ca G)}\\
                    {}\ar@{->}[r]^(.33)\delta&
                      {H^1(\ca E)}\ar@{->}[r]&
                      {H^1(\ca F)}\ar@{->}[r]&
                      {H^1(\ca G)}\\
                    {}\ar@{->}[r]^(.33)\delta&
                      {H^2(\ca E)}\ar@{->}[r]&
                      {\q\cdots\q\q}\\}
  \end{xy}
\end{displaymath}
where we have shortened $H^i(X;\ca E)$ to $H^i(\ca E)$ for the sake of conciseness and where $\delta\from H^i(X;\ca G)\to H^{i+1}(X;\ca E)$ is the so-called (Bockstein) coboundary homomorphism.

A sheaf $\ca F$ on $X$ is called acyclic if $H^i(X;L)=0$ for all $i>0$.
\begin{la}\label{constant_sheaf_is_acyclic}
Let $X$ be a partition space, $Z\subset X$ a closed subspace, and $L$ a constant
sheaf on $X$. Then $L$ is acyclic.
\end{la}
\begin{proof}
Write
\begin{displaymath}
  Z=X_{S_1}\cup\cdots\cup X_{S_r}
\end{displaymath}
with subsets $S_i\subset P$. We prove the lemma by induction on $r\in\bb N$. The case $r=0$ is trivial while that of $r=1$ follows from the fact that all open subsets of $X_S\approx X(P\sm S)$ are connected so that every constant sheaf on $X_S$ is flasque \cite{godement,hartshorne}.

In the general case $r>1$ we put
\begin{displaymath}
  Z_1=X_{S_1}\mbox{ and }Z'=X_{S_2}\cup\cdots\cup X_{S_r}
\end{displaymath}
and consider the canonical continuous surjection
\begin{displaymath}
  \tilde Z:=Z_1+Z'\stackrel{f}{\too}Z_1\cup Z'=Z
\end{displaymath}
from the disjoint sum to the union. The inverse image sheaf $f\inv L$ on $\tilde Z$ is the constant sheaf $L$ again, and the canonical homomorphism
\begin{displaymath}
  L\too f_\ast f\inv L=f_\ast L
\end{displaymath}
is injective. Its cokernel is concentrated on the intersection $Z_1\cap Z'$, and is constant there with stalk $(L\oplus L)/\Delta L$ (quotient by the diagonal). If $k\from Z_1\cap Z'\hookrightarrow Z$ denotes the inclusion we thus have obtained an exact sequence
\begin{displaymath}
  \begin{xy}
    \xymatrix{{\rule[-2.4pt]{0pt}{10pt}0}\ar@{->}[r]&
                {\rule[-2.4pt]{0pt}{10pt}L}\ar@{->}[r]&
                {f_\ast L}\ar@{->}[r]&
                {k_\ast((L\oplus L)/\Delta L)}\ar@{->}[r]&
                {\rule[-2.4pt]{0pt}{10pt}0.}\\}
  \end{xy}
\end{displaymath}
By additivity and induction the sheaf $L$ on $\tilde Z$ is acyclic, and
$k_\ast((L\oplus L)/\Delta L)$ is acyclic on $Z$, also by induction. For any subset $S\subset P$ the space $Z_1\cap f\inv U_S$ is homeomorphic to ${X(P\sm S)}_{S_1}$ or empty (if $S_1\cap S\neq\emptyset$), and $Z'\cap f\inv U_S$ to
\begin{displaymath}
  \bigcup_{\{j\,|\,S_j\cap S=\emptyset\}}{X(P\sm S)}_{S_j}.
\end{displaymath}
Once more by induction we conclude that $H^i(f\inv U_S;L)=0$ for all $i>0$, and that therefore the sheaves generated by the presheaves $U_S\mapsto H^i(f\inv U_S;L)$ also are trivial for $i>0$. This means that the Leray spectral sequence, see \cite{godement} II.4.17, for $f$ and the constant sheaf degenerates, so that
\begin{displaymath}
  H^\ast(Z;f_\ast L)=H^\ast(\tilde Z;L).
\end{displaymath}
Since $f_\ast L\to k_\ast((L\oplus L)/\Delta L)$ clearly is surjective on global
sections it follows that $L$ is acyclic on $Z$.
\end{proof}
\begin{cor}\label{qu-elem_sheaf_is_acyclic}
Let $X$ be a partition space. Then every quasi-elementary sheaf is acyclic on every closed subspace of $X$.
\end{cor}
\begin{proof}
Let $Z\subset X$ be closed, and first consider an elementary sheaf $j_!L$ on $X$, where $j\from X\sm Y\hookrightarrow X$ is the inclusion of an open subset $X\sm Y$. Writing the restricted inclusion $i\from Z\sm Y\hookrightarrow Z$ we have an exact sequence
\begin{displaymath}
  \begin{xy}
    \xymatrix{{\rule[-2.4pt]{0pt}{10pt}0}\ar@{->}[r]&
                {i_!L}\ar@{->}[r]&
                {\rule[-2.4pt]{0pt}{10pt}L}\ar@{->}[r]&
                {i_\ast L}\ar@{->}[r]&
                {\rule[-2.4pt]{0pt}{10pt}0}\\}
  \end{xy}
\end{displaymath}
of sheaves on $Z$. Since $L$ and $i_\ast L$ are acyclic by Lemma \ref{constant_sheaf_is_acyclic}, and since, $Z$ being connected, $L\to i_\ast L$ is surjective on global sections $i_!$ must also be acyclic\ts: in other words $j_!L$ is acyclic on $Z$.

The case of a general quasi-elementary sheaf clearly follows from the special case we have treated.
\end{proof}
Let $\ca F$ be a sheaf on a topological space $X$, and let $\eu U={(U_s)}_{s\in S}$ be an open covering of $X$ indexed by a stricly ordered set $S$. The so-called \v{C}ech complex is the sequence
\begin{displaymath}
  \begin{xy}
    \xymatrix{{0}\ar@{->}[r]&
                {C^0(\eu U;\ca F)}\ar@{->}[r]^\delta&
                {C^1(\eu U;\ca F)}\ar@{->}[r]^(.6)\delta&
                {\cdots}\\}
  \end{xy}
\end{displaymath}
of vector spaces
\begin{displaymath}
  C^i(\eu U;\ca F)=\prod_{s_0<\cdots<s_i}\ca F(U_{s_0}\cap\cdots\cap U_{s_i})
\end{displaymath}
and linear maps $\delta\from C^i(\eu U;\ca F)\to C^{i+1}(\eu U;\ca F)$ which are called coboundary operators and act by
\begin{displaymath}
  {(\delta x)}_{s_0\dots s_{i+1}}
    =\sum_{\alpha=0}^i{(-1)}^\alpha
     \ca F_{UV}\left(x_{s_0\dots\widehat{s_\alpha}\dots s_i}\right)
\end{displaymath}
where $V=U_{s_0}\cap\cdots\cap\widehat{U_{s_\alpha}}\cap\cdots\cap U_{s_i}$ and
$U=U_{s_\alpha}\cap V$ (by convention terms covered by a hat are to be omitted). The \v{C}ech complex indeed is a complex in the sense that the composition of any two consecutive boundary operators is $\delta\circ\delta=0$, so that for each $i$ we have inclusions
\begin{displaymath}
  B^i(\eu U;\ca F):=\im\delta\subset\ker\delta=:Z^i(\eu U;\ca F)
    \subset C^i(\eu U;\ca F).
\end{displaymath}
Elements of $C^i(\eu U;\ca F)$, $B^i(\eu U;\ca F)$, and $Z^i(\eu U;\ca F)$ are called (\v{C}ech) cochains, coboundares, and cocycles respectively, and the quotient space
\begin{displaymath}
  H^i(\eu U;\ca F):=Z^i(\eu U;\ca F)/B^i(\eu U;\ca F)
\end{displaymath}
of cocycles modulo coboundaries is the $i$-th \v{C}ech cohomology of $\ca F$ with respect to the covering $\eu U$. 

\v{C}ech cohomology does depend on the covering $\eu U$, and even if this dependence is eliminated by passing to the limit of arbitrarily fine coverings the result in general does not give the correct sheaf cohomology. For quasi-elementary sheaves things are much simpler as in fact quite a small covering is sufficient\ts:
\begin{thm}\label{may_compute_cech}
Let $X$ be a partition space and $W\subset X$ a locally closed subspace with canonical decomposition
\begin{displaymath}
  W=\bigcup_{\{(\sigma,\tau)\,|\,1\le\sigma\le r,\,1\le\tau\le s,\,
                                 S_\sigma\cap T_\tau=\emptyset\}}
    U_{S_\sigma}\cap X_{T_\tau}.
\end{displaymath}
Then for every quasi-elementary sheaf $\ca F$ on $X$ there is a canonical isomorphy
\begin{displaymath}
  H^\ast(W;\ca F)=\check{H}^\ast(\eu U;\ca F)
\end{displaymath}
between sheaf cohomology, and \v{C}ech cohomology with respect to one particular open cover $\eu U$\ts:
\begin{displaymath}
  \eu U=\Bigl(U_{S_\sigma}\cap\bigcup_{\{\tau\,|\,1\le\tau\le s,\,
                                              S_\sigma\cap T_\tau=\emptyset\}}
                          X_{T_\tau}{\Bigr)}_{\sigma=1}^r
\end{displaymath}
\end{thm}
\begin{proof}
For any $\sigma_0,\dots,\sigma_i\in\{1,\dots,r\}$ the intersection of the corresponding covering sets is
\begin{displaymath}
  U_{S_{\sigma_0}}\cap\cdots\cap U_{S_{\sigma_i}}\cap\bigcup_\tau X_{T_\tau}
    =U_{S_{\sigma_0}\cup\cdots\cup S_{\sigma_i}}\cap\bigcup_\tau X_{T_\tau}
\end{displaymath}
where the union is to be taken over those $\tau\in\{1,\dots,s\}$ for which $T_\tau$ does not meet $S_{\sigma_0}\cup\cdots\cup S_{\sigma_i}$. Thus the intersection of covering sets in question is homeomorphic to the closed subspace
\begin{displaymath}
  \bigcup_{\tau=1}^s
    {X\bigl(P\sm(S_{\sigma_0}\cup\cdots\cup S_{\sigma_i})\bigr)}_{T_\tau}
    \subset X\bigl(P\sm(S_{\sigma_0}\cup\cdots\cup S_{\sigma_i})\bigr).
\end{displaymath}
By Corollary \ref{qu-elem_sheaf_is_acyclic} the sheaf $\ca F$ is acyclic on that space, so that $\eu U$ is a Leray covering for $\ca F$. This implies the theorem by \cite{godement}~th\'eor\`eme II.5.4.1.
\end{proof}
Note that as a consequence of the theorem $H^j(W;\ca F)$ must be the zero group if $j\ge|P|$  --- in fact this is a special case of a general property of sheaves on finite dimensional topological spaces.
\subsection{Local cohomology}
\begin{figure}
  \psfrag{1}[c][c]{$1$}
  \psfrag{2}[c][c]{$2$}
  \psfrag{3}[c][c]{$3$}
  \psfrag{4}[c][c]{$4$}
  \includegraphics{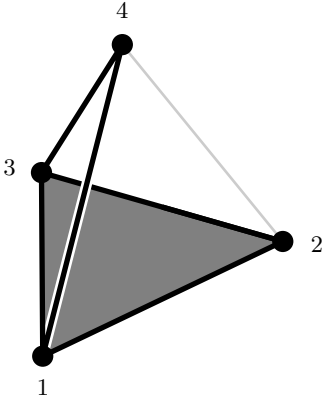}
  \caption{\label{gamma} A four party example --- the polyhedron associated to
    $Y=X_{\{1,2\}}\cup X_{\{2,3\}}\cup X_4$ is
    $\underline\Gamma_Y=\Delta_{\{3,4\}}\cup\Delta_{\{1,4\}}\cup\Delta_{\{1,2,3\}}$}
\end{figure}

The many known variants of cohomology of a sheaf $\ca F$ on a space $X$ include that of $H_\Phi^\ast(X;\ca F)$, cohomology with support in a given closed subset $\Phi\subset X$.
\begin{la}\label{local_coh_and_W}
Let $X$ be a partition space, $\Phi\subset X$ a closed subset, and $\ca F$ a
quasi-elementary sheaf on $X$. Then there is an exact sequence
\begin{displaymath}
  \begin{xy}
    \xymatrix@R0.4pc{{\rule[-2.4pt]{0pt}{10pt}0}\ar@{->}[r]&
                      {H_\Phi^0(X;\ca F)}\ar@{->}[r]&
                      {H^0(X;\ca F)}\\
                    {}\ar@{->}[r]&
                    {H^0(X\sm\Phi;\ca F)}\ar@{->}[r]^(.52)\delta&
                      {H_\Phi^1(X;\ca F)}\ar@{->}[r]&
                      {\rule[-2.4pt]{0pt}{10pt}0,}\\}
  \end{xy}
\end{displaymath}
and for every $i>1$ the coboundary homomorphism
\begin{displaymath}
  \begin{xy}
    \xymatrix{{H^{i-1}(X\sm\Phi;\ca F)}\ar@{->}[r]^(.55)\delta&
                {H_\Phi^i(X;\ca F)}\\}
  \end{xy}
\end{displaymath}
is isomorphic.
\end{la}
\begin{proof}
This follows at once from Corollary \ref{qu-elem_sheaf_is_acyclic} and the standard exact sequence
\begin{displaymath}
  \begin{xy}
    \xymatrix@R0.4pc{{\cdots}\ar@{->}[r]&
                      {H_\Phi^{i-1}(X)}\ar@{->}[r]&
                      {H^{i-1}(X)}\\
                      {}\ar@{->}[r]&
                    {H^{i-1}(X\sm\Phi)}\ar@{->}[r]^(.54)\delta&
                      {H_\Phi^i(X)}\ar@{->}[r]&
                      {\cdots}\\}
  \end{xy}
\end{displaymath}
relating cohomology with support to ordinary one.
\end{proof}
If $X=X(P)$ is a partition space with its unique closed point $\bullet$ we are
mainly interested in
\begin{displaymath}
  H_\bullet^\ast(X;\ca F):=H_{\{\bullet\}}^\ast(X;\ca F)
\end{displaymath}
which we refer to as \textit{local} cohomology.
\begin{ex}\label{local_coh_of_ext_powers}
Let $L\subset G$ be a subspace. The degree zero and one local cohomology of the exterior powers of the partition sheaf $\ca FL$ is as follows.
\begin{description}
\item[$\underline{P=\emptyset}$\q] $H_\bullet^0(X;\Lambda^k\ca FL)=\Lambda^k\{0\}$
\item[$\underline{P\neq\emptyset,\,k\!=\!0}$\q]
  $H_\bullet^0(X;\Lambda^0\ca FL)=H_\bullet^1(X;\Lambda^0\ca FL)=0$
\item[$\underline{|P|\!=\!1,\,k\!>\!0}$\q]
  $H_\bullet^0(X;\Lambda^k\ca FL)\!=\!0,\;
   H_\bullet^1(X;\Lambda^k\ca FL)\!=\!\Lambda^kL$
\item[$\underline{|P|\!>\!1,\,k\!>\!0}$\q] $H_\bullet^0(X;\Lambda^k\ca FL)
                                           =H_\bullet^1(X;\Lambda^k\ca FL)=0$
\end{description}
The last case follow from that fact that $X\sm\{\bullet\}=\bigcup_{s\in P}U_s$ is not contained in $U_p$ for any $p\in P$, so that the definition of $\ca FL$ gives $H^0(X\sm\{\bullet\};\Lambda^k\ca FL)=0$.
\end{ex}
By Lemma \ref{local_coh_and_W} calculation of local cohomology of quasi-elementary sheaves essentially comes down to the task of calculating the cohomology over the complement of the closed point. For elementary sheaves we now show how this can be explicitly done in terms of simplicial topology. We interpret the finite set $P$ as the set of vertices of a simplex $\Delta_P$ of dimension $|P|-1$\ts; thus the facets of $\Delta_P$ correspond to the non-empty subsets of $P$. A subcomplex $\underline\Gamma$ of $\Delta_P$ is given by a set $\Gamma$ of non-empty subsets of $P$ satisfying the axiom
\begin{displaymath}
  \emptyset\neq S\subset T\in\Gamma\;\Longrightarrow\;S\in\Gamma,
\end{displaymath}
and we will usually call such simplicial complexes polyhedra\ts; for more details see textbooks of topology \cite{spanier,hatcher,armstrong}. Specifically let $X=X(P)$ be a partition space, and  $Y\subset X$ a closed subspace. Then we define the \textit{associated polyhedron} $\underline\Gamma_Y$ of $Y$ by
\begin{displaymath}
  \Gamma_Y:=\left\{\emptyset\neq S\subset P\,|\,U_S\cap Y\neq\emptyset\right\}.
\end{displaymath}
In other words, to obtain $\underline\Gamma_Y$ from the simplex $\Delta_P$ one has to remove the open stars of all simplexes $S$ with $U_S$ disjoint from $Y$. In terms of the canonical decomposition
\begin{displaymath}
  Y=X_{S_1}\cup\cdots\cup X_{S_r}
\end{displaymath}
($r\in\bb N,\;S_i\subset P$ for all $i$) we have
\begin{displaymath}
  \Gamma_Y=\left\{\emptyset\neq S\subset P\,|\,S\cap S_i=\emptyset
             \mbox{ for at least one }i\right\}
\end{displaymath}
since $U_S$ meets $X_{S_i}$ if and only if $S\cap S_i$ is empty, see Fig.~\ref{gamma}.
\begin{prop}\label{hom_of_el_sheaves}
Let $X$ be a partition space, $Y\subset X$ and $Z\subset X$ closed subspaces, and $L$ a vector space. Denote by $j\from X\sm Z\hookrightarrow X$ the inclusion\ts; then there is a canonical isomorphism
\begin{displaymath}
  H^\ast(Y\sm\{\bullet\};j_!L)=H^\ast(\underline\Gamma_Y,\underline\Gamma_{Y\cap Z};L)
\end{displaymath}
between sheaf cohomology and topological cohomology with coefficient group $L$.
\end{prop}
\begin{proof}
Assume $X=X(P)$ so that $Y\sm\{\bullet\}$ is covered by the open sets
$U_p\cap Y\;(p\in P)$. We discard all the empty sets among them and apply Theorem \ref{may_compute_cech} in order to compute $H^\ast(Y\sm\{\bullet\};j_!L)$ as \v{C}ech cohomology.

To any non-empty subset $S\subset P$ the \v{C}ech complex assigns either
\begin{displaymath}
  L\mbox{,\hspace{5pt}if }U_S\cap Y\neq\emptyset\mbox{ but }U_S\cap Y\cap Z=\emptyset,
\end{displaymath}
or else the zero group. By definition the condition is met if and only if
$S\in\Gamma_Y\sm\Gamma_{Y\cap Z}$\ts: this sets up an isomorphism with the relative
simplicial cochain complex
$C^\ast(\underline\Gamma_Y)/C^\ast(\underline\Gamma_{Y\cap Z})$ of the polyhedral pair
$(\underline\Gamma_Y,\underline\Gamma_{Y\cap Z})$, and thereby proves the proposition.
\end{proof}
\begin{exs}\label{local_coh_and_polyhedra}
  $\underline\Gamma_\emptyset$ and $\underline\Gamma_{\{\bullet\}}$ are the
  empty polyhedron while
  \begin{displaymath}
    \underline\Gamma_X=\Delta_P\mbox{ and }
    \underline\Gamma_{X\setminus\{\circ\}}=\partial\Delta_P
  \end{displaymath}
  are the full simplex, respectively the simplicial sphere on the vertex set
  $P$. In particular
  \begin{itemize}
    \item in case $Y=X$ we obtain
      \begin{displaymath}
        H^\ast(X\sm\{\bullet\};j_!L)=H^\ast(\Delta_P,\underline\Gamma_Z;L),
      \end{displaymath}
      which for the particular choice $Z=X\sm\{\circ\}$ is
      \begin{displaymath}
        H^i(\Delta_P,\partial\Delta_P;L)
        =\left\lbrace
           \begin{array}{cl}
             L &\mbox{ if }i=|P|-1,\\
             0 &\mbox{ else}.
           \end{array}
        \right.
      \end{displaymath}
    \item For $Z=\emptyset$ we obtain the cohomology of the constant sheaf
      \begin{displaymath}
        H^\ast(Y\sm\{\bullet\};L)=H^\ast(\underline\Gamma_Y;L),
      \end{displaymath}
      including the special case
      \begin{displaymath}
        \begin{array}{cl}
          H^i(X\sm\{\bullet,\circ\};L)
          &=H^i(\partial\Delta_P;L)\\
        \noalign{\vskip 3pt}
          &=\left\lbrace
              \begin{array}{cl}
                L &\mbox{ if }i=0,\,|P|-2,\\
                0 &\mbox{ else}.
              \end{array}
            \right.
        \end{array}
      \end{displaymath}
    \item The long exact cohomology sequence
      \begin{displaymath}
        \begin{xy}
          \xymatrix@R0.4pc{{\cdots}\ar@{->}[r]&
                          {H^i((Y\sm\{\bullet\};j_!L)}\ar@{->}[r]&
                          {H^i((Y\sm\{\bullet\};L)}\\
                          {}\ar@{->}[r]&
                          {H^i((Y\!\cap\!Z\sm\{\bullet\};L)}
                            \ar@{->}[r]^(.59)\delta&
                          {\q\cdots\q}\\}
        \end{xy}
      \end{displaymath}
      is in perfect correspondence with that of the polyhedral pair
      \begin{displaymath}
        \begin{xy}
          \xymatrix@R0.4pc{{\cdots}\ar@{->}[r]&
                          {H^i(\underline\Gamma_Y,\underline\Gamma_{Y\cap Z})}
                            \ar@{->}[r]&
                          {H^i(\underline\Gamma_Y)}\\
                          {}\ar@{->}[r]&
                          {H^i(\underline\Gamma_{Y\cap Z})}
                            \ar@{->}[r]^\delta&
                          {\q\cdots\q}\\}
        \end{xy}
      \end{displaymath}
      as is easily verified.
  \end{itemize}
\end{exs}
\subsection{The invariants}
\label{inv_section}
Fix a set of parties $P$ and for each party $p\in P$ a finite dimensional symplectic vector space $(G_p,\omega_p)$ over the field $\bb F$. As discussed in the introduction, in the case of a prime field $\bb F$ these data determine a Heisenberg group which via the SNM representation acts on the corresponding Hilbert (state) space $\ca H$. Furthermore the stabilizer states in $H$ correspond to the isotropic subspaces $L\subset G=G_P$, a state being pure if and only if $L$ is lagrangian, that is, of half the dimension of $G$. LC equivalence of these states is the same as local symplectic equivalence of the corresponding isotropic subspaces.

To each isotropic subspace $L\subset G$ we associate the $\bb F$ vector spaces
\begin{displaymath}
  H^{ij}(L):=H_\bullet^j(X(P);\Lambda^i\ca FL)\mbox{\q for }i,j\in\bb N
\end{displaymath}
as its \textit{homological invariants}. In case of a prime field $\bb F$ we also refer to the $H^{ij}(L)$ as the homological invariants of the stabilizer states in $\ca H$ that correspond to $L$ and write $H^{ij}(|\psi\rangle)$ in the most important case where $L$ is lagrangian and that state is $|\psi\rangle\in\ca H$.
The homological invariant $H^{ij}(L)$ vanishes at least if $i>\dim L$ or if $j>|P|$, and since presently our main interest lies with $i=1$ we introduce the further abbreviation $H^j(L):=H^{1j}(L)$. While of course already the sheaf $\ca FL$ is an invariant of $L$ the homology $H^\ast(L)$ is a more compact object in which essential information on $L$ is concentrated. On the other hand $\ca FL$ has the advantage that the effect of standard operations is immediately determined.

Firstly, a given set of parties $P$ may be coarsened. This is formally described by a mapping $\phi\from P\to Q$ into another finite set $Q$. For $q\in Q$ we put
\begin{displaymath}
  H_q=G_{\{p\in P\,|\,\phi(p)=q\}}\ts;
\end{displaymath}
thus the parties from one and the same fibre of $\phi$ are combined to form a new party in $Q$ (even if $q\in Q$ is not in the image of $\phi$ it stands for a new party, though with $H_q=0$). Writing $X=X(P)$ and $Y=X(Q)$ we then have an induced continuous mapping
\begin{displaymath}
  \phi^\ast\from Y\too X;\q{\phi^\ast(y)}_p=x_{\phi(p)}.
\end{displaymath}
\begin{la}\label{coarsening}
If $L\subset G_P=G=H_Q$ is an arbitrary subspace then the corresponding partition sheaves on $X$ and $Y$, and more generally their exterior powers are related by
\begin{displaymath}
  \Lambda^\ast\ca FL={(\phi^\ast)}\inv\Lambda^\ast\ca FL.
\end{displaymath}
\end{la}
\begin{proof}
Just compute the stalks at $y\in Y$\ts:
\begin{displaymath}
  \begin{array}{cl}
    \ca FL(y) &=L\cap H_{\{q\in Q\,|\,y_q=1\}}\\
              &=L\cap G_{\{p\in P\,|\,y_{\phi(p)}=1\}}\\
              &=L\cap G_{\{p\in P\,|\,{(\phi^\ast(y))}_p=1\}}\\
              &=\ca FL(\phi^\ast(y))\\
  \end{array}
\end{displaymath}
and note that forming the inverse image sheaf commutes with stalk-wise algebraic
operations.
\end{proof}
Given a set $P$ of parties a subset $P'$ of them may be discarded --- in terms of density operators this is the process of tracing out the parties of $P'$. On the level of partition spaces we may look at the cartesian projection $\pr_P\from X(P)\to X(P\sm P')$ or alternatively at the injection $j_0\from X(P\sm P')\to X(P)$ that embeds $X(P\sm P')$ as ${X(P)}_{P'}$ in the obvious way.
\begin{la}\label{discarding}
Given $L\subset G=G_P$ let $L':=L\cap G_{P\setminus P'}$ be the space obtained by discarding the parties of $P'$, and denote by $j_0\from X(P\sm P')\to X(P)$ the injection that embeds $X(P\sm P')$ as ${X(P)}_{P'}$. Then the corresponding partition sheaf on $X(P\sm P')$ has the exterior powers
\begin{displaymath}
  \Lambda^\ast\ca FL'=j_0\inv\Lambda^\ast\ca FL.
\end{displaymath}
\end{la}
\begin{proof}
The relevant stalk $\ca FL'(x)$ at $x\in X(P\sm P')$ is
\begin{displaymath}
  L'\cap G_{\{p\in P\setminus P'\,|\,x_p=1\}}=L
    \cap G_{\{p\in P\,|\,{j_0(x)}_p=1\}}
\end{displaymath}
and thus equal to $j_0\inv\ca FL(x)$.
\end{proof}
There are two different interpretations of the tensor product of multi-party states, depending on whether the sets $P$ and $Q$ of parties that own the factors are considered disjoint (external product) or are required to be and remain identical (internal product). In the context of stabilizer states we suppose given subspaces $L\subset G_P=\bigoplus G_p$ and $M\subset H_Q=\bigoplus H_q$ in the former case, and form the family of vector spaces
\begin{displaymath}
  \left({(G\oplus H)}_r\right)_{r\in P+Q}\ts;\q
    {(G\oplus H)}_r=\left\lbrace
                      \begin{array}{cl}
                        G_r &\mbox{ if }r\in P,\\
                        H_r &\mbox{ if }r\in Q
                      \end{array}
                    \right.
\end{displaymath}
which accommodates $L\oplus M\subset{(G\oplus H)}_{P+Q}$ as the new subspace. By
contrast in the internal case ${(G_p)}_{p\in P}$ and ${(H_p)}_{p\in P}$ must be indexed by the same set of parties, and the internal direct sum $L+M$, while equal to $L\oplus M$ as an abstract space, is seen as a subspace of the total space ${(G+H)}_P$ associated with the family
\begin{displaymath}
  \bigl({(G+H)}_p\bigr)_{p\in P}\ts;\q{(G+H)}_p=G_p\oplus H_p.
\end{displaymath}
Note that the internal direct sum may be obtained from the external one by coarsening, using the diagonal map $\phi\from P\to P+P$.
\begin{la}\label{sheaves_of_products}
For the external direct sum there is a canonical graded isomorphism
\begin{displaymath}
  \Lambda^\ast\ca F(L\oplus M)
  =\Lambda^\ast\ca FL\;\hat\otimes\;\Lambda^\ast\ca FM
\end{displaymath}
of sheaves of\/ $\bb F$ algebras, the external tensor product being defined by
$\ca F\hat\otimes\ca G(x,y)=\ca F(x)\otimes\ca G(y)$ on stalks. Similarly one has
\begin{displaymath}
  \Lambda^\ast\ca F(L+M)=\Lambda^\ast\ca FL\otimes\Lambda^\ast\ca FM
\end{displaymath}
for the internal direct sum, in particular an additive isomorphism
\begin{displaymath}
  \ca F(L+M)=\ca FL\oplus\ca FM.
\end{displaymath}
\end{la}
\begin{proof}
The general statements are clear and imply the last since
\begin{displaymath}
\Lambda^1\ca F(L+M)
  =\Lambda^0\ca FL\!\otimes\!\Lambda^1\ca FM
   \,\oplus\,\Lambda^1\ca FL\!\otimes\!\Lambda^0\ca FM
\end{displaymath}
and $\Lambda^0\ca FL=\Lambda^0\ca FM=\bb F$.
\end{proof}
In general the functorial behaviour of partition sheaves does not determine the effect of coarsening, discarding, or internal products on the homological invariants, with the exception of
\begin{displaymath}
  H^j(L+M)=H^j(L)\oplus H^j(M)\mbox{\q for all }j\in\bb N
\end{displaymath}
which is an obvious corollary to Lemma \ref{sheaves_of_products}. A tensor product formula in homology also holds for the external product\ts: this will be discussed in the next section.
\section{Products}
\label{prod_section}
\subsection{External tensor product}
\begin{thm}\label{product_thm}
Let $X=X(P)$ and $Y=X(Q)$ be partition spaces, and assume that $\ca F$ and $\ca G$ are quasi-elementary sheaves on $X$ respectively $Y$. Then the external tensor product $\ca F\,\hat\otimes\,\ca G$, taken over the base field $\bb F$, is a quasi-elementary sheaf on $X\times Y=X(P+Q)$. The cohomology cross product
\begin{displaymath}
  H_\bullet^\ast(X;\ca F)\otimes H_\bullet^\ast(Y;\ca G)
  \stackrel{\times}{\too}
  H_\bullet^\ast(X\times Y;\ca F\,\hat\otimes\,\ca G)
\end{displaymath}
is an isomorphism.
\end{thm}
\begin{proof}
It is clear that the (internal or external) tensor product of elementary sheaves is elementary. The analogous statement for quasi-elementary sheaves follows from this since tensor products over a field preserve exact sequences.

Consider a short exact sequence
\begin{displaymath}
  \begin{xy}
    \xymatrix{{0}\ar@{->}[r]&
                {\ca G'}\ar@{->}[r]&
                {\ca G}\ar@{->}[r]&
                {\ca G''}\ar@{->}[r]&
                {0}\\}
  \end{xy}
\end{displaymath}
of quasi-elementary sheaves on $Y$. Abbreviating $H=H_\bullet^\ast(X;\ca F)$ we have a commutative diagram
\begin{displaymath}
  \begin{xy}
    \xymatrix{{\vdots}\ar@{->}[d]^(.45)\delta&
                {\vdots}\ar@{->}[d]^(.45)\delta\\
              {H\otimes H_\bullet^\ast(Y;\ca G')}
                \ar@{->}[r]^(.45)\times\ar@{->}[d]&
              {H_\bullet^\ast(X\times Y;\ca F\,\hat\otimes\,\ca G')}
                \ar@{->}[d]\\
              {H\otimes H_\bullet^\ast(Y;\ca G)}
                \ar@{->}[r]^(.45)\times\ar@{->}[d]&
              {H_\bullet^\ast(X\times Y;\ca F\,\hat\otimes\,\ca G)}
                \ar@{->}[d]\\
              {H\otimes H_\bullet^\ast(Y;\ca G'')}
                \ar@{->}[r]^(.45)\times\ar@{->}[d]^(.4)\delta&
              {H_\bullet^\ast(X\times Y;\ca F\,\hat\otimes\,\ca G'')}
                \ar@{->}[d]^(.4)\delta&\\
              {\vdots}&
                {\vdots}\\}
  \end{xy}
\end{displaymath}
with exact columns. Thus if the cross product is isomorphic for $\ca G'$ and $\ca G''$ in place of $\ca G$ it must be so for $\ca G$ too, in view of the five lemma for exact sequences. The proof of the theorem is thus reduced to the case of an elementary sheaf $\ca G$ by the obvious induction. By another application of the same argument we may assume that $\ca F$ is elementary as well.

Our strategy will be to express the local cohomology groups like $H_\bullet^\ast(X)$ in terms of non-local cohomology over the punctured spaces $A:=X\sm\{\bullet\}$ and $B:=Y\sm\{\bullet\}$, using Lemma \ref{local_coh_and_W}. We claim that this is compatible with the cross product in the sense that for all $i,j\in\bb N$ the diagram shown as Fig.~\ref{xs_and_deltas} is commutative up to sign, even for arbitrary sheaves $\ca F$ and $\ca G$.
\begin{figure*}
  \begin{displaymath}
    \begin{xy}
      \xymatrix{{H_\bullet^i(X;\ca F)\otimes H_\bullet^j(Y;\ca G)}
                  \ar@{->}[rr]^\times&&
                {H_\bullet^{i+j}(X\times Y;\ca F\,\hat\otimes\,\ca G)}\\
                {H_\bullet^i(X;\ca F)\otimes H^{j-1}(B;\ca G)}
                  \ar@{->}[r]^\times
                  \ar@{->}[u]^{\id\otimes\,\delta''}&
                {H_{\bullet\times B}^{i+j-1}
                (X\times B;\ca F\,\hat\otimes\,\ca G)}
                  \ar@{->}[ur]^{\delta''}&&
                {H^{i+j-1}(A\by Y\cup X\by B;\ca F\,\hat\otimes\,\ca G)}
                  \ar@{->}[ul]_\delta\\
                {H^{i-1}(A;\ca F)\otimes H^{j-1}(B;\ca G)}
                  \ar@{->}[rr]^\times
                  \ar@{->}[u]^{\delta'\otimes\,\id}&&
                {H^{i+j-2}(A\times B;\ca F\,\hat\otimes\,\ca G)}
                  \ar@{->}[ul]^{\delta'}
                  \ar@{->}[ur]_{\delta_{\rm MV}}&\\}
    \end{xy}
  \end{displaymath}
  \caption{\label{xs_and_deltas} Compatibility of the cross product with
    coboundaries}
\end{figure*}
Indeed commutativity in the left half of the diagram is clear. The diamond on the right hand side involves a Mayer-Vietoris coboundary, and in terms of a flasque resolution of $\ca F\,\hat\otimes\,\ca G$ both paths from bottom to top are represented as follows on the cochain level. Given a class
$\xi\in H^{i+j-2}(A\times B)$
\begin{itemize}
  \item represent it by a cochain $x$ over $X\times Y$ (which must be a cocycle
    over $A\times B$), and
\item pick a $(i+j-1)$-cochain $y$ over $X\times Y$ with
  \begin{displaymath}
    y|(A\times B)=0\mbox{ and }x|(X\times B)=y|(X\times B)\ts;
  \end{displaymath}
\end{itemize}
then $\delta y$ represents, up to sign, both $\delta''\delta'\xi$ and
$\delta\delta_{\rm MV}\xi$ in $H_\bullet^{i+j}(X\times Y)$.

We now do assume that $\ca F=I_!F$ and $\ca G=J_!G$ are elementary, say with respect to the open inclusions
\begin{displaymath}
  I\from X\sm W\hookrightarrow X\mbox{ and }J\from Y\sm Z\hookrightarrow Y.
\end{displaymath}
For the moment we require that the closed subsets $W$ and $Z$ both strictly contain $\{\bullet\}$. Lemma \ref{local_coh_and_W} implies that then
\begin{displaymath}
  \begin{array}{ll}
    H_\bullet^0(X;I_!F) &=H^0(X;I_!F)\\
                        &=H^0(A;I_!F)= H_\bullet^1(X;I_!F)=0,\\
  \end{array}
\end{displaymath}
so that for all (even non-positive) $i\in\bb Z$ the coboundary homomorphism
$H^{i-1}(A;\ca F)\stackrel{\delta}{\too}H_\bullet^i(X;\ca F)$ is bijective. Analogous statements hold for the sheaves $\ca G$ and
$\ca F\,\hat\otimes\,\ca G$.

A look at Fig.~\ref{xs_and_deltas} shows that in order to prove the theorem it suffices to show that the homomorphism of degree $+1$
\begin{displaymath}
  \begin{array}{ll}
    H^\ast(A;\ca F)\otimes H^\ast(B;\ca G)
      &\stackrel{\times}{\too}
        H^\ast(A\times B;\ca F\,\hat\otimes\,\ca G)\\
      &\stackrel{\delta_{\rm MV}}{\too}
        H^\ast(A\by Y\cup X\by B;\ca F\,\hat\otimes\,\ca G)
  \end{array}
\end{displaymath}
is bijective. But now Proposition \ref{hom_of_el_sheaves} provides a translation to the simplicial setting, turning the homomorphism in question into
\begin{displaymath}
  \begin{array}{ll}
    &H^\ast(\Delta_P,\underline\Gamma_W;F)\otimes
     H^\ast(\Delta_Q,\underline\Gamma_Z;G)\\
    \stackrel{\times}{\too}
    &H^\ast(\Delta_P\by\Delta_Q,
       \underline\Gamma_W\by
       \Delta_Q\cup\Delta_P\by\underline\Gamma_Z;F\otimes G)\\
       \stackrel{\delta_{\rm MV}}{\too}
    &H^\ast(\Delta_P\ast\Delta_Q,
            \underline\Gamma_W\ast\Delta_Q\cup\Delta_P\ast\underline\Gamma_Z;
            F\otimes G).
  \end{array}
\end{displaymath}
Indeed we have
$\ca F\,\hat\otimes\,\ca G=I_!F\,\hat\otimes\,J_!G={(I\by J)}_!(F\otimes G)$
with the inclusion
\begin{displaymath}
  I\by J\from(X\by Y)\sm(W\by Y\cup X\by Z)\hookrightarrow X\by Y,
\end{displaymath}
and the corresponding polyhedra are
\begin{displaymath}
  \underline\Gamma_{X\times Y}=\underline\Gamma_X\ast\underline\Gamma_Y
    =\Delta_P\ast\Delta_Q,
\end{displaymath}
 respectively the polyhedron $\underline\Gamma_{W\by Y\cup X\by Z}$ whose set of simplexes is
\begin{displaymath}
  \left\{\emptyset\neq S\!+\!T\subset P\!+\!Q\,|\,
    U_S\cap W\neq\emptyset\mbox{ or }U_T\cap Z\neq\emptyset\right\}.
\end{displaymath}
We have used the asterisk to indicate the join
$\underline\Gamma\ast\underline\Theta$ of two polyhedra $\underline\Gamma$ and
$\underline\Theta$. Geometrically speaking, the join puts a closed interval between every point of $\underline\Gamma$ and every point of $\underline\Theta$ --- but the combinatorial definition is even simpler and makes
\begin{displaymath}
  \{\emptyset\neq S\!+\!T\,|\,
    S=\emptyset\mbox{ or }S\in\Gamma,\,
    \mbox{ and }T=\emptyset\mbox{ or }T\in\Theta\}
\end{displaymath}
(set sum, that is \textit{disjoint} union of $S$ and $T$) the set of simplexes of $\underline\Gamma\ast\underline\Theta$.

Since $W$ and $Z$ strictly contain $\{\bullet\}$ the polyhedra $\underline\Gamma_W$ and 
$\underline\Gamma_Z$ are non-empty, and therefore
\begin{displaymath}
  \underline\Gamma_{W\by Y\cup X\by Z}
    =\underline\Gamma_W\ast\Delta_Q\cup\Delta_P\ast\underline\Gamma_Z
\end{displaymath}
Finally the simplicial version of $\delta_{\rm MV}$ is the Mayer-Vietoris coboundary obtained from bisection of the connecting coordinate of the join. This $\delta_{\rm MV}$ clearly is an isomorphism, and so ist the topological cross product since the tensor product is taken over a field. This proves isomorphy under the extra assumptions on  $W$ and $Z$.

In the more special cases $W=\{\bullet\}$ or $Z=\{\bullet\}$ we still have isomorphisms $H^{i-1}(A;\ca F)\stackrel{\delta}{\too}H_\bullet^i(X;\ca F)$ etc.\ for all $i\in\bb Z$, and the only modification needed is that now $\underline\Gamma_W$ or $\underline\Gamma_Z$ may be empty, so that $\underline\Gamma_{W\by Y\cup X\by Z}$ becomes one of
\begin{displaymath}
  \emptyset\!+\!\Delta_Q\cup\Delta_P\ast\underline\Gamma_Z,\;
  \underline\Gamma_W\ast\Delta_Q\cup\Delta_P\!+\!\emptyset\mbox{, or }
  \Delta_P\!+\!\Delta_Q.
\end{displaymath}

The remaining cases, say with $W=\emptyset$, are essentially trivial. We have $\Gamma_W=\emptyset$ and furthermore $H_\bullet^\ast(X;\ca F)=H_\bullet^\ast(X;F)=0$. In case $Z\neq\emptyset$ the local cohomology of $X\times Y$ is
\begin{displaymath}
  H_\bullet^\ast(X\times Y;F\hat\otimes\ca G)
    =H^\ast(X\by Z;F\hat\otimes\ca G),
\end{displaymath}
and in view of $\underline\Gamma_{X\times Z}=\Delta_P\ast\underline\Gamma_Z$ and the topological fact $H^\ast(\Delta_P\ast\Delta_Q,\Delta_P\ast\underline\Gamma_Z)=0$ this is the trivial group. If also $Z=\emptyset$ then $\ca F\hat\otimes\ca G=F\!\otimes\!G$ is a constant sheaf and $H_\bullet^\ast(X\by Y;F\!\otimes\!G)=0$. Thus there is nothing to prove in any of these cases.
\end{proof}

Combining Theorem \ref{product_thm} with Lemma \ref{sheaves_of_products} we immediately obtain\ts:
\begin{thm}\label{hom_product}
Assume that a base field $\bb F$ and families of finite dimensional vector spaces ${(G_p)}_{p\in P}$ and ${(H_q)}_{q\in Q}$ over $\bb F$ are given, as well as vector subspaces $L\subset G=G_P$ and $M\subset H=H_Q$. The cross product then defines an isomorphism
\begin{displaymath}
  \begin{array}{rcl}
    H_\bullet^\ast(X;\Lambda^\ast\ca FL)
      &\otimes &H_\bullet^\ast(Y;\Lambda^\ast\ca FM)\\
  \noalign{\vskip 3pt}
      &&\stackrel{\simeq}{\too}
        H_\bullet^\ast(X\times Y;\Lambda^\ast\ca F(L\oplus M))
  \end{array}
\end{displaymath}
of bigraded $\bb F$ algebras.
\end{thm}
\begin{cor}\label{cohom_of_reducible}
If the subspace $L\subset G$ is a direct sum $L=L_1\oplus\cdots\oplus L_r$
for some partition $P=P_1+\cdots+P_r$ of the set of parties then
\begin{displaymath}
  H_\bullet^\ast(X;\Lambda^j\ca FL)=0\mbox{\q for }j=0,\dots,r\!-\!1.
\end{displaymath}
In particular the full partition sheaf $\ca G:=\ca FG$ has trivial cohomology
$H_\bullet^\ast(X;\ca G)=0$ as soon as $|P|\ge2$.
\end{cor}
\begin{proof}
$\Lambda^0\ca FL=\bb F$ is the constant sheaf, and therefore
$H_\bullet^\ast(X;\Lambda^0\ca FL)=0$. Now Theorem \ref{hom_product} gives the result.
\end{proof}
\subsection{Duality}
In this section we fix a partition space $X=X(P)$, a corresponding collection ${(G_p)}_{p\in P}$ of finite dimensional vector spaces over the base field $\bb F$ and, furthermore for each $p\in P$ a bilinear form
\begin{displaymath}
  \omega_p\from G_p\otimes G_p\too\bb F.
\end{displaymath}
Thus
\begin{displaymath}
  \omega:=\sum_{p\in P}\omega_p
\end{displaymath}
is a bilinear form on $G$, and the direct sum decomposition $G=\bigoplus_{p\in P}G_p$ is orthogonal with respect to $\omega$.

We form the closed subspace
\begin{displaymath}
  Z:=\bigcup_{S\subset P}X_S\by X_{P\setminus S}\subset X\times X
\end{displaymath}
and let
\begin{displaymath}
  k\from(X\by X)\sm Z=\bigcup_{p\in P}U_p\by U_p\hookrightarrow X\times X
\end{displaymath}
be the inclusion of the complement.

\begin{figure}
  \psfrag{[;01]}[rb][rb]{$[;01]$}
  \psfrag{[01;]}[lt][lt]{$[01;]$}
  \psfrag{[0;1]}[rb][rb]{$[0;1]$}
  \psfrag{[1;0]}[lb][lb]{$[1;0]$}
  \includegraphics{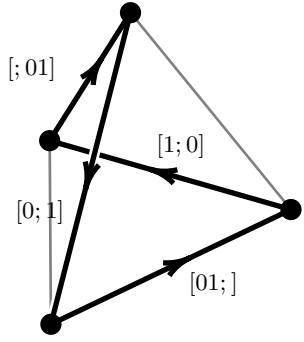}
  \caption{\label{top_sphere} The oriented sphere $\underline\Gamma_Z$ for $P=\{0,1\}$}
\end{figure}

Let now $L\subset G$ and $M\subset G$ be linear subspaces that are orthogonal in the sense that $\omega(L\otimes M)=0$. Writing $\ca G=\ca FG$ as before we have a homomorphism of sheaves on $X\times X$
\begin{displaymath}
  \ca G/\ca FL\;\;\hat\otimes\;\;\ca FM\stackrel{\hat\omega}{\too}k_!\bb F
\end{displaymath}
which on the presheaf level, say over the open set $U\by V\subset X\by X$ with $S=\{p\,|\,U\subset U_p\}$, $T=\{p\,|\,V\subset U_p\}$ is given by
\begin{displaymath}
  \begin{array}{ccccc}
    G_S/(L\cap G_S) &\otimes &(M\cap G_T) &\too    &\bb F\\
    \overline x     &\otimes &y           &\mapsto &\omega(x,y).
  \end{array}
\end{displaymath}
Note that $\omega(x,y)$ can be non-zero only if the supports of $x$ and $y$ intersect, which implies $U\subset U_p$ and $V\subset U_p$ simultaneously for some $p\in P$, whence $(U\by V)\cap Z=\emptyset$\ts: thus $\hat\omega$ indeed takes values in $k_!\bb F$.
\begin{la}\label{sphere}
The subpolyhedron $\underline\Gamma_Z\subset\Delta_{P+P}=\Delta_P\ast\Delta_P$ is given by
\begin{displaymath}
  \Gamma_Z=\left\{\emptyset\neq S+T\,\subset\,P+P\,|\,S\cap T=\emptyset\right\}
\end{displaymath}
and is a topological sphere of dimension $|P|\!-\!1$. If an orientation of $\Delta_P$ is chosen, say by ordering $P=\{0,1,\dots,l\!-\!1\}$ then $\underline\Gamma_Z$ is oriented by its ordered faces
\begin{displaymath}
  {(-1)}^r{(-1)}^{\pi(s,t)}[s_1\dots s_r;t_1\dots t_{l-r}]
\end{displaymath}
with support $\{s_1,\dots,s_r\}+\{t_1,\dots,t_{l-r}\}\subset P+P$. Here $\pi(s,t)$ stands for the permutation with ordered values $s_1,\dots,s_r,t_1,\dots,t_{l-r}$, see Fig.~\ref{top_sphere}.
\end{la}
\begin{proof}
By definition we have
\begin{displaymath}
  \begin{array}{rcl}
    \Gamma_Z &= &\left\{\emptyset\neq S+T\subset P+P\,
                   \left|\,\begin{array}{l}
                             (S+T)\cap(R+P\sm R)\\
                             =\emptyset\mbox{ for some }R\subset P
                           \end{array}
                   \right.
                 \right\}\\
  \noalign{\vskip 3pt}
             &= &\{\emptyset\neq S+T\subset P+P\,|\,S\cap T=\emptyset\}.
  \end{array}
\end{displaymath}
By induction on $|P|\ge0$ we prove that $\underline\Gamma_Z$ is a sphere. For
$P=\emptyset$ clearly $\underline\Gamma_Z=\emptyset$ is a $(-1)$-sphere. On the other hand let $\tilde Z\subset\tilde X$ denote the subset of the partition space $\tilde X=\{0,1\}\times X$ associated to the set of parties $\{p\}+P$. Then
\begin{displaymath}
  \begin{array}{rcl}
    \Gamma_{\tilde Z} &=    &\big\{\{p\}+\emptyset\big\}\\
  \noalign{\vskip 3pt}
                      &\cup &\big\{(\{p\}\cup S)+T\,|\,S+T\in\Gamma_Z\big\}\\
  \noalign{\vskip 3pt}
                      &\cup &\Gamma_Z\\
  \noalign{\vskip 3pt}
                      &\cup &\big\{S+(\{p\}\cup T)\,|\,S+T\in\Gamma_Z\big\}\\
  \noalign{\vskip 3pt}
                      &\cup &\big\{\emptyset+\{p\}\big\}
  \end{array}
\end{displaymath}
so that $\underline\Gamma_{\tilde Z}$ is the (unreduced) suspension \cite{spanier,hatcher,armstrong} of $\underline\Gamma_Z$. Since $\underline\Gamma_Z$ is a sphere by induction hypothesis $\underline\Gamma_{\tilde Z}$ too is a sphere of the correct dimension.

The statement on orientation follows by direct calculation.
\end{proof}
\begin{cor}\label{sphere_evaluation}
Assuming $|P|\ge2$ we have isomorphisms
\begin{displaymath}
  \begin{array}{rcl}
    H_\bullet^j(X\by X;k_!\bb F)
      &=      &H^{j-1}((X\by X)\sm\{\bullet\};k_!\bb F)\\
  \noalign{\vskip 3pt}
      &=      &H^{j-1}(\Delta_{P+P},\underline\Gamma_Z;\bb F)\\
  \noalign{\vskip 3pt}
      &\simeq &\left\{\begin{array}{ll}
                        \bb F &\mbox{\rm for }j=|P|+1,\\
                        0     &\mbox{\rm else,}
                      \end{array}
               \right.
  \end{array}
\end{displaymath}
which are canonical apart from the last one which depends on the choice of an orientation for $\Delta_P$. If $p\in P$ is an arbitrarily chosen party then this isomorphism may be realised as evaluation on the fundamental class $\underline\Delta_p$ given by
\begin{displaymath}
  \Delta_p=\left\{\emptyset\neq S+T\,\subset\,P+P\,\big|\,S\cap T\subset\{p\}\right\}
\end{displaymath}
with face orientations
\begin{displaymath}
  {(-1)}^p{(-1)}^{\pi(s,t)}[ps_1\dots s_{r-1};t_1\dots t_{|P|-r}].
\end{displaymath}
\end{cor}
\begin{proof}
Combine Lemma \ref{local_coh_and_W}, Proposition \ref{hom_of_el_sheaves}, and Lemma \ref{sphere} and calculate for the fundamental class.
\end{proof}
\begin{figure*}
  \begin{displaymath}
    \begin{xy}
      \xymatrix{{\cdots}\ar@{->}[r]
                  &{H_\bullet^\ast(\ca FL/\ca FM)}
                    \ar@{->}[r]\ar@{->}[d]^{\hat\omega}
                  &{H_\bullet^\ast(\ca G/\ca FM)}
                    \ar@{->}[r]\ar@{->}[d]^{\hat\omega}
                  &{H_\bullet^\ast(\ca G/\ca FL)}
                    \ar@{->}[r]\ar@{->}[d]^{\hat\omega}
                  &{\cdots}\\
                {\cdots}\ar@{->}[r]
                  &{\Hom\big(H_\bullet^\ast(\ca FM^\perp\!/\ca FL^\perp),
                     \bb F\big)}
                    \ar@{->}[r]
                  &{\Hom\big(H_\bullet^\ast(\ca FM^\perp),\bb F\big)}
                    \ar@{->}[r]
                  &{\Hom\big(H_\bullet^\ast(\ca FL^\perp),\bb F\big)}
                    \ar@{->}[r]
                  &{\cdots}\\
                  }
    \end{xy}
  \end{displaymath}
  \caption{\label{five_lemma} Duality morphisms for a pair of subspaces $M\subset L\subset G$}
\end{figure*}
Returning to the situation before Lemma \ref{sphere} we assume $|P|\ge2$ and fix an orientation of $\Delta_P$. The bilinear pairing on the sheaf level $\hat\omega\from\ca G/\ca FL\;\;\hat\otimes\;\;\ca FM\too k_!\bb F$ induces one of cohomology
\begin{displaymath}
  H_\bullet^\ast(X\times X;\ca G/\ca FL\;\hat\otimes\;\ca FM)\too
    H_\bullet^\ast(X\times X;k_!\bb F).
\end{displaymath}
Composing with the isomorphisms
\begin{displaymath}
  \begin{array}{rl}
    H_\bullet^\ast(X;\ca FL)
      &\otimes\hspace{6pt}H_\bullet^\ast(X;\ca FM)\\
  \noalign{\vskip 3pt}
      &\stackrel{\simeq}{\too}
          H_\bullet^\ast(X;\ca G/\ca FL)\otimes H_\bullet^\ast(X;\ca FM)\\
  \noalign{\vskip 3pt}
      &\stackrel{\simeq}{\too}
          H_\bullet^\ast(X\times X;\ca G/\ca FL\;\hat\otimes\;\ca FM)\\
  \end{array}
\end{displaymath}
and
\begin{displaymath}
  \begin{array}{rl}
    H_\bullet^\ast(X&\!\!\!\times\,X;k_!\bb F)\\
  \noalign{\vskip 3pt}
      &\stackrel{\simeq}{\too}
          H^\ast((X\times X)\sm\{\bullet\};k_!\bb F)\\
  \noalign{\vskip 3pt}
      &\stackrel{\simeq}{\too}
          H^\ast(\Delta_{P+P},\underline\Gamma_Z;k_!\bb F)\\
  \noalign{\vskip 3pt}
      &\stackrel{\simeq}{\too}
          H^{|P|}(\Delta_{P+P},\underline\Gamma_Z;k_!\bb F)\\
  \noalign{\vskip 3pt}
      &\stackrel{\simeq}{\too}
          \bb F\\
  \end{array}
\end{displaymath}
we obtain an equivalent pairing
\begin{displaymath}
  H_\bullet^\ast(X;\ca FL)\otimes H_\bullet^\ast(X;\ca FM)\too\bb F
\end{displaymath}
which we refer to as the \textit{pairing induced by} $\omega$ and likewise write $\hat\omega$. By construction it is homogeneous of degree $|P|+2$, that is, it non-trivially pairs $H_\bullet^i(X;\ca FL)$ with $H_\bullet^j(X;\ca FM)$ only if $i+j=|P|+2$.
\begin{thm}\label{duality_thm}
We assume that $|P|\ge2$ and that for each $p\in P$ the bilinear form $\omega_p\from G_p\otimes G_p\to\bb F$ is a perfect pairing of vector spaces. Then for every linear subspace $L\subset G$ the induced pairing
\begin{displaymath}
  H_\bullet^\ast(X;\ca FL)\otimes H_\bullet^\ast(X;\ca FL^\perp)
    \stackrel{\hat\omega}{\too}\bb F
\end{displaymath}
also is perfect, where the orthogonal complement refers to $\omega\from G\otimes G\to\bb F$.
\end{thm}
\begin{proof}
We will eventually prove the equivalent statement that $H_\bullet^\ast(X;\ca G/\ca FL)\otimes H_\bullet^\ast(X;\ca FL^\perp)\to\bb F$ is a perfect pairing. As a first step we observe the following generalisation of the duality pairing. Let $M\subset L$ be a second subspace, then
\begin{displaymath}
  \omega\from L/M\otimes M^\perp/L^\perp\too\bb F
\end{displaymath}
is perfect, and we obtain an induced bilinear form
\begin{displaymath}
  H_\bullet^\ast(X;\ca FL/\ca FM)\otimes
  H_\bullet^\ast(X;\ca FM^\perp\!/\ca FL^\perp)
  \too\bb F
\end{displaymath}
as before. The exact sequences
\begin{displaymath}
  \begin{xy}
    \xymatrix@R0.4pc{{\rule[-2.4pt]{0pt}{10pt}0}\ar@{->}[r]
                &{\ca FM}\ar@{->}[r]
                &{\ca FL}\ar@{->}[r]
                &{\ca FL/\ca FM}\ar@{->}[r]
                &{\rule[-2.4pt]{0pt}{10pt}0}\\
              {\rule[-2.4pt]{0pt}{10pt}0}\ar@{->}[r]
                &{\ca FL^\perp}\ar@{->}[r]
                &{\ca FM^\perp}\ar@{->}[r]
                &{\ca FM^\perp/\ca FL^\perp}\ar@{->}[r]
                &{\rule[-2.4pt]{0pt}{10pt}0}\\}
  \end{xy}
\end{displaymath}
and the duality pairings induce the commutative diagram of vector spaces with exact rows shown as Fig.~\ref{five_lemma}. If two out of any two vertical arrows are bijective then so are all, by the five lemma. Since given $L$ a subspace $M\subset L$ may be picked arbitrarily we may perform induction on $\dim L$. This reduces the question to the claim that for any $M\subset L$ of codimension one the pairing
\begin{displaymath}
  H_\bullet^\ast(X;\ca FL/\ca FM)\otimes H_\bullet^\ast(X;\ca FM^\perp/\ca FL^\perp)
  \stackrel{\hat\omega}{\too}\bb F
\end{displaymath}
is perfect.

\textit{Topological digression.} Fix a finite set $P\neq\emptyset$, and consider a proper subpolyhedron $\underline\Gamma$ of the simplex $\Delta_P$. Then the set
\begin{displaymath}
  \Gamma\tschech:=\left\{\emptyset\neq S\subset P\,|\,
    \emptyset\neq P\sm S\notin\Gamma\right\}
\end{displaymath}
defines another proper subpolyhedron $\underline\Gamma\tschech\subset\Delta_P$ which is the \textit{dual polyhedron} of $\underline\Gamma$ (in $\Delta_P$), see
Fig.~\ref{gamma_dual}. It is at once seen to obey the rules
\begin{itemize}
\item $\underline\Gamma_1\subset\underline\Gamma_2\;\;\Rightarrow\;\;
       \underline\Gamma_1\tschech\supset\underline\Gamma_2\tschech$,
\item ${(\underline\Gamma_1\cup\underline\Gamma_2)}\tschech=
         \underline\Gamma_1\tschech\cap\underline\Gamma_2\tschech,\q
      {(\underline\Gamma_1\cap\underline\Gamma_2)}\tschech=
         \underline\Gamma_1\tschech\cup\underline\Gamma_2\tschech$,
\item ${\underline\Gamma\tschech}\tschech=\underline\Gamma$.
\end{itemize}
From Lemma \ref{sphere} we already know that the subpolyhedron
$\underline\Lambda\subset\Delta_P\ast\Delta_P=\Delta_{P+P}$ given by
\begin{displaymath}
  \Lambda=\left\{\emptyset\neq S+T\,\subset\,P+P\,|\,S\cap T=\emptyset\right\}
\end{displaymath}
is a $(|P|\!-\!1)$-dimensional sphere.
\begin{la}\label{top_inclusion}
$\underline\Lambda\;\subset\;\underline\Gamma\tschech\ast\Delta_P\;\cup\;
\Delta_P\ast\underline\Gamma$ holds for every proper $\underline\Gamma\subset\Delta_P$.
\end{la}
\begin{proof}
Let $S+T\in\Lambda$ be arbitrary. If $S=\emptyset$ or $T=\emptyset$ then certainly
$S+T$ is a simplex of $\underline\Gamma\tschech\ast\Delta_P$ respectively
$\Delta_P\ast\underline\Gamma$. Thus assume $S\neq\emptyset\neq T$, and furthermore
that $T\notin\Gamma$. But then
\begin{displaymath}
  \emptyset\neq T\subset P\sm S\notin\Gamma
\end{displaymath}
shows that $S\in\Gamma\tschech$ and therefore $S+T\in\Gamma\tschech+P$.
\end{proof}
The lemma allows to write down the topological duality pairing
\begin{displaymath}
  \begin{array}{rcl}
    H^\ast(\Delta_P,\underline\Gamma\tschech)
      &\otimes &H^\ast(\Delta_P,\underline\Gamma)\\
  \noalign{\vskip 3pt}
      &&\stackrel{\times}{\too}
          H^\ast(\Delta_P\by\Delta_P,
            \Gamma\tschech\by\Delta_P\cup\Delta_P\by\underline\Gamma)\\
  \noalign{\vskip 3pt}
      &&\stackrel{\delta_{\rm MV}}{\too}
          H^\ast(\Delta_P\ast\Delta_P,
            \Gamma\tschech\ast\Delta_P\cup\Delta_P\ast\underline\Gamma)\\
  \noalign{\vskip 3pt}
      &&\too
          H^\ast(\Delta_P\ast\Delta_P,\underline\Lambda)\\
  \noalign{\vskip 3pt}
      &&\;\simeq\;\;\bb F\\
  \end{array}
\end{displaymath}
(of topological cohomology with an arbitrary coefficient ring $\bb F$). It is a pairing of degree $|P|$  and depends on the choice of an orientation for $\Delta_P$ as usual.
\begin{figure}
  \psfrag{1}[c][c]{$1$}
  \psfrag{2}[c][c]{$2$}
  \psfrag{3}[c][c]{$3$}
  \psfrag{4}[c][c]{$4$}
  \includegraphics{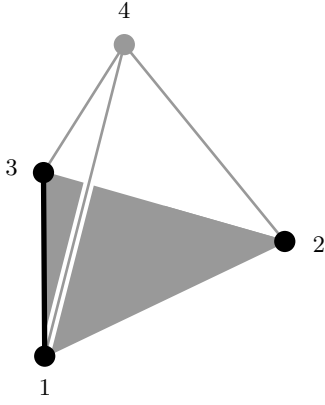}
  \caption{\label{gamma_dual} $\Delta_{\{1,3\}}\cup\Delta_{\{2\}}$ is the dual of the
    polyhedron $\underline\Gamma_Y$ of Fig.~\ref{gamma}}
\end{figure}
\begin{la}\label{poincare}
This pairing is perfect.
\end{la}
\begin{proof}
This is a version of Poincar\'e duality, and a proof is indicated but for the sake of convenience. The statement is true if $\Gamma=\emptyset$ since in that case the pairing becomes
\begin{displaymath}
  \begin{array}{rcl}
    H^\ast(\Delta_P)
      &\otimes &H^\ast(\Delta_P,\partial\Delta_P)\\
  \noalign{\vskip 3pt}
      &&\stackrel{\times}{\too}
          H^\ast(\Delta_P\by\Delta_P,
            \Delta_P\by\partial\Delta_P)\\
  \noalign{\vskip 3pt}
      &&\stackrel{\delta}{\too}
          H^\ast(\Delta_P\ast\Delta_P,
            \emptyset\ast\Delta_P\cup\Delta_P\ast\partial\Delta_P)\\
  \noalign{\vskip 3pt}
      &&\too
          H^\ast(\Delta_P\ast\Delta_P,\underline\Lambda)\\
  \noalign{\vskip 3pt}
      &&\;\simeq\;\;\bb F\\
  \end{array}
\end{displaymath}
where each arrow is an isomorphism. The statement also holds trivially in the case where $\Gamma=\{p\}$ consists of a single vertex, since then $\Gamma\tschech=\{\emptyset\neq S\subset P\,|\,P\sm\{p\}\neq S\neq P\}$, so that both $\underline\Gamma$ and $\underline\Gamma\tschech$ are contractible.

The general case now follows by induction on the number of simplexes in $\underline\Gamma$, using Mayer-Vietoris sequences and naturality of the duality pairing.
\end{proof}
We return to the proof of Theorem \ref{duality_thm} and assume that $M\subset L$ is a subspace of codimension one. Composing the isomorphisms from Lemma \ref{local_coh_and_W} and Proposition \ref{hom_of_el_sheaves} we obtain
\begin{displaymath}
  \begin{array}{rcl}
    H_\bullet^\ast(X;\ca FL/\ca FM)
      &\simeq &H^\ast(W;\ca FL/\ca FM)\\
  \noalign{\vskip 3pt}
      &\simeq &H^\ast(W;i_!L/M)\\
  \noalign{\vskip 3pt}
      &\simeq &H^\ast(\Delta_P,\underline\Gamma;L/M)\\
  \end{array}
\end{displaymath}
where $W=X\sm\{\bullet\}$, where
\begin{displaymath}
  i\from\bigcup_{\{\emptyset\neq S\subset P\,|\,L\cap H_S\not\subset M\}}U_S
    \hookrightarrow W
\end{displaymath}
is the open inclusion, and $\underline\Gamma\subset\Delta_P$ the corresponding polyhedron\ts: $\Gamma=\left\{\emptyset\neq S\subset P\,|\,L\cap H_S\subset M\right\}$. We have an analogous isomorphism
\begin{displaymath}
  \begin{array}{rcl}
    H_\bullet^\ast(X;\ca FM^\perp\!/\ca FL^\perp)
      &\simeq &H^\ast(W;\ca FM^\perp\!/\ca FL^\perp)\\
  \noalign{\vskip 3pt}
      &\simeq &H^\ast(W;i_!M^\perp\!/L^\perp)\\
  \noalign{\vskip 3pt}
      &\simeq &H^\ast(\Delta_P,\underline\Gamma';M^\perp\!/L^\perp),\\
  \end{array}
\end{displaymath}
where the subpolyhedron $\underline\Gamma'\subset\Delta_P$ is given by
\begin{displaymath}
  \Gamma'=\left\{\emptyset\neq S\subset P\,|\,
            M^\perp\cap H_S\subset L^\perp\right\}.
\end{displaymath}
\begin{la}\label{L_M_perp}
For every subset $S\subset P$ the following three statements are equivalent.
\begin{enumerate}
  \renewcommand{\labelenumi}{\rm(\arabic{enumi})}
  \item $M^\perp\cap H_S\subset L^\perp$,
  \item $L+H_{P\setminus S}=M+H_{P\setminus S}$,
  \item $L\cap H_{P\setminus S}\not\subset M$.
\end{enumerate}
In particular $\underline\Gamma'=\underline\Gamma\tschech$ is the dual polyhedron.
\end{la}
\begin{proof}
$(1)\!\Leftrightarrow(2)$\ts: property (2) means $L\subset M+H_{P\setminus S}$, and since $\omega$ is perfect we may translate this to
\begin{displaymath}
  M^\perp\cap H_S={(M+H_{P\setminus S})}^\perp\subset L^\perp,
\end{displaymath}
which is just (1).

$(2)\!\Rightarrow\!(3)$\ts: pick any $g\in L\sm M$. By assumption we may write
\begin{displaymath}
  g=f+h\mbox{\q with }f\in M,\;h\in H_{P\setminus S}.
\end{displaymath}
Then $g-f\in L\cap H_{P\setminus S}\sm M$, which shows (2).

$(3)\Rightarrow(2)$\ts: choose any $g\in L\cap H_{P\setminus S}\sm M$, then
\begin{displaymath}
  L=M+\langle g\rangle\subset M+H_{P\setminus S},
\end{displaymath}
and (3) follows.

We finally verify that
\begin{displaymath}
  \begin{array}{rcl}
    \Gamma\tschech
      &= &\{\emptyset\neq S\subset P\,|\,P\sm S\neq\emptyset\mbox{ and }
            L\cap H_{P\setminus S}\not\subset M\}\\
  \noalign{\vskip 3pt}
      &= &\{\emptyset\neq S\subset P\,|\,P\sm S\neq\emptyset\mbox{ and }
            M^\perp\cap H_S\subset L^\perp\}\\
  \noalign{\vskip 3pt}
      &= &\{S\in\Gamma'\,|\,P\sm S\neq\emptyset\}\\
  \noalign{\vskip 3pt}
      &= &\Gamma',\\
  \end{array}
\end{displaymath}
using that $P\notin\Gamma'$ since $M^\perp\neq L^\perp$ anyway.
\end{proof}

We can now easily complete the proof of Theorem \ref{duality_thm}. The pairing in question is 
\begin{displaymath}
  H_\bullet^\ast(X;\ca FL/\ca FM)\otimes
    H_\bullet^\ast(X;\ca FM^\perp/\ca FL^\perp)
    \stackrel{\hat\omega}{\too}\bb F
\end{displaymath}
and we have just seen how to translate sheaf cohomology to topological cohomology, which results in a pairing
\begin{displaymath}
  H^\ast(\Delta_P,\underline\Gamma;L/M)\otimes
    H^\ast(\Delta_P,\underline\Gamma\tschech;M^\perp/L^\perp)
    \too\bb F.
\end{displaymath}
This is easily seen to coincide up to a degree dependent sign with the Poincar\'e duality pairing of Lemma \ref{poincare}, whence it is perfect.
\end{proof}
\subsection{A duality formula}
It is useful to be able to describe the duality pairing explicitly in terms of \v{C}ech cohomology. To this end we may return to the more general situation where $L$ and $M$ are arbitrary linear subspaces of $G$ that are orthogonal with respect to $\omega$. As before we write $\ca G=\ca FG=\ca FG_P$\ts; also recall the notations $W=X\sm\{\bullet\}$, and $k$ for the inclusion of $\bigcup_{p\in P}U_p\times U_p$ in $X$, as well as $\Lambda=\{\emptyset\neq S+T\subset P+P\,|\,S\cap T=\emptyset\}$. The extensive diagram which involves the relevant mappings is displayed as Fig.~\ref{pairing_diag}.
\begin{figure*}
  \begin{displaymath}
    \begin{xy}
      \xymatrix{{H^\ast(W;\ca FL)\otimes H^\ast(W;\ca FM)}
                     \ar@{->}[r]_\simeq^{\delta\otimes\delta}
                  &{H_\bullet^\ast(X;\ca FL)\otimes H_\bullet^\ast(X;\ca FM)}\\
                {H^\ast(W;\ca G/\ca FL)\otimes H^\ast(W;\ca FM)}                   
                     \ar@{->}[r]_\simeq^{\delta\otimes\delta}
                     \ar@{->}[u]_\simeq^{\delta\otimes\id}
                     \ar@{->}[d]_\times
                  &{H_\bullet^\ast(X;\ca G/\ca FL)\otimes
                    H_\bullet^\ast(X;\ca FM)}
                     \ar@{->}[u]_{\delta\otimes\id}^\simeq
                     \ar@{->}[dd]^\times\\
                {H^\ast(W\by W;\ca G/\ca FL\;\hat\otimes\;\ca FM)}
                     \ar@{->}[d]_{\delta_{\rm MV}}\\
                {H^\ast((X\by X)\sm\{\bullet\};
                  \ca G/\ca FL\;\hat\otimes\;\ca FM)}
                     \ar@{->}[r]_\simeq^\delta
                     \ar@{->}[d]_\omega
                  &{H_\bullet^\ast(X\by X;\ca G/\ca FL\;\hat\otimes\;\ca FM)}
                     \ar@{->}[d]^\omega\\
                {H^\ast((X\by X)\sm\{\bullet\};k_!\bb F)}
                     \ar@{->}[r]_\simeq^\delta
                     \ar@{=}[d]
                  &{H_\bullet^\ast(X\by X;k_!\bb F)}
                     \ar@{->}[d]\\
                {H^\ast(\Delta_{P+P},\underline\Lambda;\bb F)}
                     \ar@{->}[r]
                  &{\bb F}\\}
    \end{xy}
  \end{displaymath}
  \caption{\label{pairing_diag} The diagram comparing two versions of the duality pairing commutes up to a degree dependent sign}
\end{figure*}
Take two cohomology classes $x\in H^i(W;\ca FL)$ and $y\in H^j(W;\ca FM)$ with
$i+j=l:=|P|$, and represent them by \v{C}ech cocycles
\begin{displaymath}
  (x_{s_0s_1\dots s_i})\mbox{ and }(y_{t_0t_1\dots t_j}),
\end{displaymath}
each index $s=s_0s_1\dots s_i$ and $t=t_0t_1\dots t_j$ representing an ordered simplex of dimension $i$ and $j$ respectively. Let us write both $x$ and $y$ as coboundaries
\begin{displaymath}
  x=\delta u\mbox{ and }y=\delta v
\end{displaymath}
with values in $\ca G$, that is, explicitly\ts:
\begin{displaymath}
  \begin{array}{rcl}
    x_{s_0\dots s_i}
      &= &\sum_{\alpha=0}^i{(-1)}^\alpha
           u_{s_0\dots\widehat{s_\alpha}\dots s_i}\\
  \noalign{\vskip 3pt}
    y_{t_0\dots t_j}
      &= &\sum_{\beta=0}^j{(-1)}^\beta v_{t_0\dots\widehat{t_\beta}\dots t_j}\\
  \end{array}
\end{displaymath}
We fix an arbitrary $p\in P$ and realise the isomorphism $H^l(\Delta_{P+P},\underline\Lambda;\bb F)\simeq\bb F$ as evaluation on the oriented polyhedron $\underline\Delta_p$. From the commutativity of the diagram Fig.~\ref{pairing_diag} we conclude that up to a fixed sign the value $\hat\omega(x,y)$ of the duality pairing is
\begin{displaymath}
  \sum_{\begin{array}{rcl}
          \scriptstyle S+T &\scriptstyle= &\scriptstyle P\setminus\{p\}\\
        \noalign{\vskip -3pt}
          \scriptstyle |S| &\scriptstyle= &\scriptstyle i-1\\
        \noalign{\vskip -3pt}
          \scriptstyle |T| &\scriptstyle= &\scriptstyle j
        \end{array}}
    {(-1)}^p{(-1)}^{\pi(s,t)}\omega(u_{ps_1\dots s_{i-1}},y_{pt_1\dots t_j})
\end{displaymath}
where it is understood that for each set $S$ an oriented simplex $[s_1\dots s_{i-1}]$ with support $S$ is chosen, and similarly for $T$. Substituting the expression of $y$ as a coboundary and noting that $u_{ps_1\dots s_{i-1}}$ and $v_{t_1\dots t_j}$ have disjoint supports as vectors in $G$ we obtain
\begin{displaymath}
  \begin{array}{rl}
    \displaystyle\sum_{S+T=P\setminus\{p\}}\!\!
      &{(-1)}^p{(-1)}^{\pi(s,t)}\cdot\\
      &\cdot\sum_{\beta=1}^j{(-1)}^\beta
         \omega(u_{ps_1\dots s_{i-1}},v_{pt_1\dots\widehat{t_\beta}\dots t_j}).
  \end{array}
\end{displaymath}
Replacing $T$ by the smaller set $T\sm\{t_\beta\}$ we rewrite this expression as
\begin{displaymath}
    \displaystyle\sum_{\begin{array}{rcl}
                         \scriptstyle S+T &\scriptstyle\subset &\scriptstyle P\setminus\{p\}\\
                       \noalign{\vskip -3pt}
                         \scriptstyle |S| &\scriptstyle= &\scriptstyle i-1\\
                       \noalign{\vskip -3pt}
                         \scriptstyle |T| &\scriptstyle= &\scriptstyle j-1
                       \end{array}}\!\!\!\!
      {(-1)}^p{(-1)}^{\pi(s,t)}\epsilon\,
  \omega(u_{ps_1\dots s_{i-1}},v_{pt_1\dots t_{j-1}})
\end{displaymath}
with $\epsilon={(-1)}^\beta{(-1)}^{\pi(p,t_\beta)}{(-1)}^{i+t_\beta-\beta}$ if $t_\beta$ is the unique element of $P\sm(S+T+\{p\})$. We finally obtain\ts:
\begin{prop}\label{formula}
Let $L,M\subset G$ be linear subspaces with $\omega(L\otimes M)=0$, and let $x\in H^i(W,\ca FL)$ and $y\in H^j(W;\ca FM)$, with $i+j=|P|$. Pick $\ca G$-valued cochains $u,v$ with $\delta u=x$ and $\delta v=y$. Then
\begin{displaymath}
  \begin{array}{rl}
    \hat\omega(x,y)
      &=\displaystyle\sum_{\begin{array}{rcl}
                             \scriptstyle S+T &\scriptstyle\subset
                               &\scriptstyle P\setminus\{p\}\\
                           \noalign{\vskip -3pt}
                             \scriptstyle |S| &\scriptstyle=
                               &\scriptstyle i-1\\
                           \noalign{\vskip -3pt}
                             \scriptstyle |T| &\scriptstyle=
                               &\scriptstyle j-1
                           \end{array}}\!\!\!\!
        {(-1)}^{p+i+r}{(-1)}^{\pi(p,r)}{(-1)}^{\pi(s,t)}\cdot\\
      &\q\q\q\q\q\q\q\q\cdot\omega_p(u_{ps_1\dots s_{i-1}},v_{pt_1\dots t_{j-1}})
  \end{array}
\end{displaymath}
if $r\in P$ denotes the unique element not in $S+T+\{p\}$.
\end{prop}
An essential aspect of the formula of Proposition \ref{formula} is that it treats both factors in $\hat\omega(x,y)$ on an equal footing. Thus $\hat\omega$ might as well have been defined using $\ca FL$ and $\ca G/\ca FM$ in the same way as we have used $\ca G/\ca FL$ and $\ca FM$. Apart from being interesting in its own right this fact has further consequences if $\omega$ enjoys symmetry properties. For the sake of simplicity we state but the case of characteristic two.
\begin{thm}\label{symmetry_thm}
Assume that the characteristic of the base field $\bb F$ is two, and that for each $p\in P$ the bilinear form $\omega_p$ on $G_p$ is alternating. If the linear subspace $L\subset G$ is  isotropic for $\omega$ then
\begin{displaymath}
  \hat\omega\from H_\bullet^\ast(X;\ca FL)\otimes H_\bullet^\ast(X;\ca FL)\too\bb F
\end{displaymath}
also alternates.
\end{thm}
\begin{proof}
The formula of Proposition \ref{formula} shows that for homogeneous $x,y\in H_\bullet^\ast(X;\ca FL)$ we have $\hat\omega(x,y)+\hat\omega(y,x)=0$\ts: just swap $S$ and $T$. Therefore it suffices to know that $\hat\omega(x,x)=0$ for all $x\in H_\bullet^i(X;\ca FL)$ with $2i=|P|+2$. But this also follows from the formula.
\end{proof}
\begin{cor}\label{symplectic}
Assume $\chr\bb F=2$, that all $\omega_p$ are symplectic forms, and that $L\subset G$ is a lagrangian subspace\ts: $L=L^\perp$. Then
\begin{displaymath}
  \hat\omega\from H_\bullet^\ast(X;\ca FL)\otimes H_\bullet^\ast(X;\ca FL)\too\bb F
\end{displaymath}
is symplectic. If $|P|$ is even then $\hat\omega$ restricts to a symplectic form on $H_\bullet^{|P|/2-1}(X;\ca FL)$.
\end{cor}
\begin{proof}
Combine Theorem \ref{duality_thm} and Theorem \ref{symmetry_thm}.
\end{proof}
\section{Examples and applications}
\label{ex_appl_section}
\begin{table*}
  \begin{tabular}{|c|c|c||c|c|c|c|c|c||c|c|c|c|c|c||c|c|c|c|c|c||c|c|c|c|c|c|}
    \hline
    \rule[0pt]{0pt}{10pt}\bf Graph &\bf Name &$l$
      &$h^{10}$ &$h^{11}$ &$h^{12}$ &$h^{13}$ &$h^{14}$ &$h^{15}$ &$h^{20}$ &$h^{21}$ &$h^{22}$ &$h^{23}$ &$h^{24}$ &$h^{25}$ &$h^{30}$ &$h^{31}$ &$h^{32}$ &$h^{33}$ &$h^{34}$ &$h^{35}$ &$h^{40}$ &$h^{41}$ &$h^{42}$ &$h^{43}$ &$h^{44}$ &$h^{45}$ \\
    \hline\hline
No.\ke1 &$|A_2\rangle$ &2 &0 &0 &2 & & & & & & & & & & & & & & & & & & & & & \\
    \hline
No.\ke2 &$|A_3\rangle$ &3 &0 &0 &1 &1 & & &0 &0 &0 &3 & & & & & & & & & & & & & & \\
    \hline
No.\ke3 &$|\mbox{GHZ}_4\rangle$ &4 &0 &0 &1 &0 &1 & &0 &0 &0 &1 &3 & &0 &0 &0 &0 &4 & & & & & & & \\
    \hline
No.\ke4 &$|\hat A_3\rangle$ &4 &0 &0 &0 &2 &0 & &0 &0 &0 &1 &3 & &0 &0 &0 &0 &4 & & & & & & & \\
    \hline
No.\ke5 &$|\mbox{GHZ}_5\rangle$ &5 &0 &0 &1 &0 &0 &1 &0 &0 &0 &1 &0 &4 &0 &0 &0 &0 &1 &6 &0 &0 &0 &0 &0 &5 \\
    \hline
No.\ke6 &$|D_5\rangle$ &5 &0 &0 &0 &1 &1 &0 &0 &0 &0 &0 &3 &2 &0 &0 &0 &0 &1 &6 &0 &0 &0 &0 &0 &5 \\
    \hline
No.\ke7 &$|A_5\rangle$ &5 &0 &0 &0 &1 &1 &0 &0 &0 &0 &0 &4 &1 &0 &0 &0 &0 &1 &6 &0 &0 &0 &0 &0 &5 \\
    \hline
No.\ke8 &$|\hat A_4\rangle$ &5 &0 &0 &0 &1 &1 &0 &0 &0 &0 &0 &5 &0 &0 &0 &0 &0 &1 &6 &0 &0 &0 &0 &0 &5 \\    \hline
  \end{tabular}
  \caption{\label{five}
    Homological invariants of graph states with up to five parties. Graphs are
    labelled as in \cite{hein3} and an alternative standard name for each graph
    or the corresponding state is shown. $l=|P|$ is the number of parties, and
    $h^{ij}$ shorthand for $\dim_{\bb F_2}H^{ij}(L)$. For reasons of space the
    trivial information $h^{lj}=0$ for $j<l$ and $h^{ll}=1$ is not shown.}
\end{table*}
\begin{table}
  \begin{tabular}{|c|c|c||c|c|c|c|c|c|c|c|}
    \hline
    \rule[0pt]{0pt}{10pt}\bf Graph &\bf Name &$l$
      &$h^0$ &$h^1$ &$h^2$ &$h^3$ &$h^4$ &$h^5$ &$h^6$ &$h^7$ \\
    \hline\hline
    No.\ke9 &$|\mbox{GHZ}_6\rangle$ &6 &0 &0 &1 &0 &0 &0 &1 & \\
    \hline
    No.\ke10 & &6 &0 &0 &0 &1 &0 &1 &0 & \\
    \hline
    No.\ke11 &$|\hat D_5\rangle$ &6 &0 &0 &0 &0 &2 &0 &0 & \\
    \hline
    No.\ke12 &$|D_6\rangle$ &6 &0 &0 &0 &1 &0 &1 &0 & \\
    \hline
    No.\ke13 &$|E_6\rangle$ &6 &0 &0 &0 &0 &2 &0 &0 & \\
    \hline
    No.\ke14 &$|A_6\rangle$ &6 &0 &0 &0 &0 &2 &0 &0 & \\
    \hline
    No.\ke15 & &6 &0 &0 &0 &1 &0 &1 &0 & \\
    \hline
    No.\ke16 & &6 &0 &0 &0 &0 &6 &0 &0 & \\
    \hline
    No.\ke17 & &6 &0 &0 &0 &0 &2 &0 &0 & \\
    \hline
    No.\ke18 &$|\hat A_5\rangle$ &6 &0 &0 &0 &0 &4 &0 &0 & \\
    \hline
    No.\ke19 & &6 &0 &0 &0 &0 &12 &0 &0 & \\
    \hline
    No.\ke20 &$|\mbox{GHZ}_7\rangle$ &7 &0 &0 &1 &0 &0 &0 &0 &1 \\
    \hline
    No.\ke21 & &7 &0 &0 &0 &1 &0 &0 &1 &0 \\
    \hline
    No.\ke22 & &7 &0 &0 &0 &0 &1 &1 &0 &0 \\
    \hline
    No.\ke23 & &7 &0 &0 &0 &1 &0 &0 &1 &0 \\
    \hline
    No.\ke24 &$|\hat D_6\rangle$ &7 &0 &0 &0 &1 &0 &0 &1 &0 \\
    \hline
    No.\ke25 & &7 &0 &0 &0 &0 &1 &1 &0 &0 \\
    \hline
    No.\ke26 & &7 &0 &0 &0 &0 &1 &1 &0 &0 \\
    \hline
    No.\ke27 &$|D_7\rangle$ &7 &0 &0 &0 &0 &1 &1 &0 &0 \\
    \hline
    No.\ke28 &$|E_7\rangle$ &7 &0 &0 &0 &0 &1 &1 &0 &0 \\
    \hline
    No.\ke29 &$|\hat E_6\rangle$ &7 &0 &0 &0 &0 &1 &1 &0 &0 \\
    \hline
    No.\ke30 &$|A_7\rangle$ &7 &0 &0 &0 &0 &1 &1 &0 &0 \\
    \hline
    No.\ke31 & &7 &0 &0 &0 &1 &0 &0 &1 &0 \\
    \hline
    No.\ke32 & &7 &0 &0 &0 &0 &3 &3 &0 &0 \\
    \hline
    No.\ke33 & &7 &0 &0 &0 &0 &1 &1 &0 &0 \\
    \hline
    No.\ke34 & &7 &0 &0 &0 &0 &1 &1 &0 &0 \\
    \hline
    No.\ke35 & &7 &0 &0 &0 &0 &1 &1 &0 &0 \\
    \hline
    No.\ke36 & &7 &0 &0 &0 &0 &3 &3 &0 &0 \\
    \hline
    No.\ke37 & &7 &0 &0 &0 &0 &1 &1 &0 &0 \\
    \hline
    No.\ke38 & &7 &0 &0 &0 &0 &2 &2 &0 &0 \\
    \hline
    No.\ke39 & &7 &0 &0 &0 &0 &1 &1 &0 &0 \\
    \hline
    No.\ke40 &$|\hat A_6\rangle$ &7 &0 &0 &0 &0 &1 &1 &0 &0 \\
    \hline
    No.\ke41 & &7 &0 &0 &0 &0 &2 &2 &0 &0 \\
    \hline
    No.\ke42 & &7 &0 &0 &0 &0 &1 &1 &0 &0 \\
    \hline
    No.\ke43 & &7 &0 &0 &0 &0 &3 &3 &0 &0 \\
    \hline
    No.\ke44 & &7 &0 &0 &0 &0 &1 &1 &0 &0 \\
    \hline
    No.\ke45 & &7 &0 &0 &0 &0 &6 &6 &0 &0 \\
    \hline
  \end{tabular}
  \caption{\label{six_sev}
    First order homological invariants of graph states with six or seven
    parties, with $h^j=\dim_{\bb F_2}H^j(L)$}
\end{table}
In order to illustrate  homological invariants we have calculated, using \cite{singular}, a variety of examples including all pure stabilizer states in up to seven parties controlling a single qubit each. The base field here is $\bb F_2$, and for each $p\in P$ one has $G_p=\bb F_2^2=\bb F_2e\oplus\bb F_2f$ with the standard symplectic form\ts: $\omega_p(e,f)=1$. As is well known under these assumptions all lagrangian subspaces of $G=G_P$ are equivalent, under local symplectic transformations, to standard forms that may be described by simple graphs on the vertex set $P$. Though this description does not play an immediate role in the present context we make use of the classification in \cite{hein3} of lagrangians in terms of graphs, retaining the labelling used in that work for the sake of convenient reference. We show the complete result for up to five parties in Table~\ref{five}, and just the information on $H^j(L)=H^{1j}(L)$ in the case of six or seven parties in Table~\ref{six_sev}.

Note that $H^{0j}(L)=0$ for all $j$ as soon as $P$ is non-empty, and that the homological invariants of any tensor products of the listed states may be calculated by Theorem \ref{hom_product}. In particular for products with more than one factor --- those which define decomposable lagrangians and stabilizer states --- the first order invariants $H^j(L)=H^{1j}(L)$ vanish by Corollary \ref{cohom_of_reducible}.

In the case of parties controlling just one qubit every indecomposable state automatically is irreducible in the sense that it cannot, in a non-trivial way, be written as a tensor product of stabilizer states distributed over the same set of parties. In terms of lagrangians this notion of irreducibility means that there is no way to write the lagrangian as an internal direct sum of smaller lagrangians. In view of this the fact that $H^1(L)$ vanishes in all cases of the lists is no coincidence but a consequence of\ts:
\begin{prop}\label{h1_of_irr}
Let $L\subset G=\bigoplus_{p\in P}G_p$ be an isotropic subspace and assume that $L':=L\cap G_p$ is non-zero for some party $p\in P$. Then $L$ splits off a summand in $G_p$\ts: there is an orthogonal splitting $G=G'\oplus G''$ with $0\neq G'\subset G_p$ such that $L=L'\oplus L''$ with $L'':=L\cap G''$.
\end{prop}
\begin{proof}
Since $L'\subset G_p$ is isotropic we may pick a subspace $G'\subset G_p$ of dimension $2\cdot\dim L'$ which contains $L'$ and to which $\omega_p$ restricts as a symplectic form. We define $G''\subset G$ as the orthogonal complement of $G'$ in $G$ and have $G=G'\oplus G''$ as claimed. The inclusion $L'\oplus L''\subset L$ is clear, and equality follows from a comparison of dimensions\ts:
\begin{displaymath}
  \begin{array}{rcl}
    \dim L'' &=   &\dim{(L^\perp+G')}^\perp\\
             &=   &\dim G-\dim L^\perp-\dim G'+\dim(L^\perp\cap G')\\
             &\ge &\dim L-\dim G'+\dim(L\cap G')\\
             &=   &\dim L-\dim G'+\dim L'\\
  \end{array}
\end{displaymath}
so that $\dim L'+\dim L''\ge\dim L-\dim G'+2\dim L'=\dim L$.
\end{proof}
The first order invariants $H^j(L)$ for $j\ge2$ shown in the tables nicely illustrate the duality properties described in Theorem \ref{duality_thm} and Corollary \ref{symplectic}.

We do not know whether there exist undecomposable stabilizer states with trivial first order invariants but such states must involve at least four parties. We first make a technical but more general statement and give a proof which is adapted from that of \cite{bravyi3} Theorem 5.
\begin{prop}\label{reduce_trivial_hom}
Let $L\subset G=\bigoplus_{p\in P}G_p$ be a lagrangian subspace with $L\cap G_p=0$ for all $p\in P$. Assume that for some subset $P'\subset P$ we have
$H^2(L\cap G_{P'})\neq0$ and
\begin{displaymath}
  (L\cap G_{P'})+\sum_{\{Q\subset P\,|\,P'\not\subset Q\}}(L\cap P_Q)=L.
\end{displaymath}
Then there exist a subspace $G'\subset G_{P'}$ such that $\omega$ restricts to a symplectic form on $G'$ and such that $L':=L\cap G'$ is a lagrangian subspace of $G'$ which defines a GHZ state. $G$ and $L$ split as internal direct sums $G=G'\oplus G''$ and $L=L'\oplus L''$ with $G'':=G^\perp$ and $L'':=L\cap G''$.
\end{prop}
\begin{proof}
Any cocycle
\begin{displaymath}
 (g_{st})\in H^1(W;\ca F(L\cap G_{P'}))=H^2(L\cap G_{P'})
\end{displaymath}
may be represented as a coboundary $g_{st}=f_s-f_t$ with values in $\ca FG_{P'}$, and since $L\cap G_p=0$ for all $p\in P$ we have a well-defined bilinear form
\begin{displaymath}
  \hat\omega\from H^2(L\cap G_{P'})\otimes L\too\bb F\ts;\q
  \hat\omega(g_{st},h)=\omega(f_s,h).
\end{displaymath}
We claim that if $\hat\omega(g_{st},h)=0$ for all $h\in L\cap G_{P'}$ then $(g_{st})=0$. Indeed, if $Q\subset P$ is a subset with $P'\not\subset Q$, say with $p\in P'\sm Q$ then trivially $\hat\omega(g_{st},h)=\omega(f_p,h)=0$. Thus if $\hat\omega(g_{st},h)=0$ holds for all $h\in L\cap G_{P'}$ then it even holds for all $h\in L$, in view of the hypothesis of the proposition. Since $L$ is lagrangian it follows that $f_s\in L\cap G_s$, and therefore that $f_s$ and a fortiori all $g_{st}$ vanish.

Now pick any cocycle $(g_{st}=f_s-f_t)$ and let $h\in L\cap G_{P'}$ be an element such that $\hat\omega(g_{st},h)=1$. For each $s\in P'$ let $G_s'\subset G_s$ be the subspace spanned by $f_s$ and the component of $h$ in $G_s$\ts: thus each $G_s'$ is a symplectic plane. Putting $G'=\bigoplus_{s\in P'}G_s'$ the intersection $L\cap G'$ is spanned by the $f_s$ and $h$, and the conclusion of the proposition is readily verified.
\end{proof}
\begin{cor}\label{three_parties}
Let $P$ be a set of three parties and let $L\subset G=\bigoplus_{p\in P}G_p$ be a lagrangian subspace with $H^\ast(L)=0$ Then $L$ is decomposable.
\end{cor}
\begin{proof}
If $L\cap G_p$ is non-zero for some $p$ then $L$ splits of a summand in $G_p$, by Proposition \ref{h1_of_irr}. We thus may assume $L\cap G_p=0$ for all $p$. If for some two-party set $P'\subset P$ we have $H^2(L\cap G_{P'})\neq0$ we may apply Proposition \ref{reduce_trivial_hom} since the assumption $H^3(L)=0$ implies $\sum_{|Q|=2}(L\cap P_Q)=L$. Thus in that case $L$ splits off an EPR state. There remains the case where $H^2(L\cap G_{P'})=0$ for all $P'\subset P$ with $|P|=2$. Explicitly this condition means $L\cap G_{P'}=0$ for all such $P'$\ts; since $H^3(L)=0$ on the other hand means that the coboundary homomorphism $\bigoplus_{|P'|=2}L\cap G_{P'}\to L$ is surjective we conclude that $L=0$ which of course is dcomposable too.
\end{proof}
One of the main results of \cite{bravyi3} is a formula for for the number of GHZ states that can be extracted from a multi-party stabilizer state by LC operations. We here state and reprove this result in terms of homological invariants.
\begin{thm}\label{ghz_extraction}
Let $L\subset G=\bigoplus_{p\in P}G_p$ be a lagrangian subspace. The number of lagrangians of all-party GHZ states that can be split off from $L$ is
\begin{displaymath}
  \frac{1}{2}\left(\dim_\bb FH^2(L)+\dim_\bb FH^{|P|}(L)\right)
\end{displaymath}
(which coincides with $\dim H^2(L)=\dim H^{|P|}(L)$ as soon as $|P|>2$).
\end{thm}
\begin{proof}
It suffices to show that such a lagrangian can be split off if and only if $H^2(L)\neq0$. We put $l=|P|$. The condition is necessary by Lemma \ref{sheaves_of_products} since the GHZ state has the non-trivial first order invariants $H^2(L)$ and $H^l(L)$, each of dimension one unless $l=2$ when they merge in a single space of dimension two.

Now assume that $H^2(L)$ is non-trivial. Put $X=X(P)$, $W=X\sm\{\bullet\}$, and $\ca G=\ca FG$ as usual. By Theorem \ref{duality_thm} the pairing $\hat\omega\from H_\bullet^2(X;\ca G/\ca FL)\otimes H_\bullet^l(X;\ca FL)\to\bb F$ is perfect, so we may pick cocycles
\begin{displaymath}
  (f_s)\in H^0(W;\ca G/\ca FL)=H_\bullet^1(X;\ca G/\ca FL)=H_\bullet^2(X;\ca FL)\end{displaymath}
and
\begin{displaymath}
  h_P\in H^{l-1}(W;\ca FL)=H_\bullet^l(X;\ca FL)
\end{displaymath}
with $\hat\omega(f_s,h_P)=1$. Represent $h_P$ by any vector $h={(h_s)}_{s\in P}\in L$. Then for each $s\in P$ the vectors $f_s$ and $h_s$ span a symplectic plane in $G_s$, and $L\cap G'$ is spanned by the  $f_s$ and $h$, and is a GHZ lagrangian.
\end{proof}
The authors of \cite{bravyi3} construct an example of an irreducible four-party stabilizer state $|\psi\rangle$ which is not a GHZ state. This property can easily be read from the first order homological invariants of $|\psi\rangle$, which have dimensions $h^2\!=\!0$, $h^3\!=\!4$, and $h^4\!=\!0$, as follows\ts: Two of the four parties own one qubit, and the others two qubits each. If the lagrangian of $|\psi\rangle$ were an internal direct sum, one summand would have to involve all parties since $h^3\!\neq\!0$, and the complementary summand would have trivial invariants. But by the classification indecomposable four-party states in four or five qubits with $h^3\!=\!4$ do not exist.

Scanning the classification lists we found many more six- and seven-party states that by suitable coarsening lead to a four-party state with $h^3\!=\!4$\ts; they include the states No.\ke14, 17, 18, 30, 33, 35, and 37--45. Amusingly, the six-party state No.19 from which \cite{bravyi3}'s example is constructed in fact gives a four-party state with $h^3\!=\!4$ whichever coarsening is chosen, as long as two of the new parties are given two qubits each.
\section{Conclusion}
\label{concl_section}
In this paper we have introduced homological invariants of a system of vector spaces which is partitioned over a finite set of parties, and thereby LC invariants of multi-party stabilizer states. We have investigated some basic properties of these invariants, in particular their duality. In the simplest cases we have explicitly calculated the invariants, and we have shown their connection with known results on the extraction of GHZ states from stabilizer states.

We believe that the potential of homological invariants in fact reaches much further, and wish to suggest several ways in which we believe the present work can be continued.

As mentioned in the introduction LC equivalence of stabilizer seems to be little stricter than LU equivalence, the conceptually more important notion. While homological invariants refer to LC equivalence by definition it is nevertheless possible that they in fact are LU invariants. The most satisfactory positive answer to that question would, of course, involve an extension of our construction from stabilizer to all quantum states.

Homological invariants do not, nor are intended to separate all the different LC classes of states. Yet it may be true, for an arbitrary number of parties, that the first order invariant $H^\ast(|\psi\rangle)$ vanishes \textit{only} if $|\psi\rangle$ is decomposable. Quite generally it would be worthwhile to obtain a global view of the homological classification of states, including asymptotic information on the size of the equivalence classes for large sets of parties, or with respect to large (but finite) field extensions. For instance this might give a way to re-interpret and extend the results of \cite{smith2} which show that stabilizer states from which many GHZ states can be extracted --- that is, those with large $H^2(|\psi\rangle)$ --- are exceptional rather than the rule.

Our main result, the Duality Theorem \ref{duality_thm}, has so far been used in quite a limited way, relating $H^2$ and $H^{|P|}$. It would be reasonable to expect that a more systematic application will lead to an improved qualitative understanding of multi-party entanglement of stabilizer states, completing the picture for up to four or five parties at least. Systematic use of homological invariants probably will take advantage of the fact that rather than being mere numbers they are algebraic structures with a well-defined functorial behaviour, some of which we have explained in Section \ref{inv_section}. This pertains even more to the higher order invariants $H^{ij}(L)$ --- so far unused at all ---which combine to form the algebra $H^\ast(L)=\bigoplus_{ij}H^{ij}(L)$ and thus carry a multiplication as an additional structure.

\bibliography{hom_inv}

\end{document}